\documentclass{article}
\usepackage{amsthm}

\usepackage{mathrsfs}
\usepackage{graphicx}
\usepackage{algorithm}
\usepackage{algpseudocode}
\usepackage[preprint]{neurips_2025}

\usepackage{titletoc}

\usepackage[utf8]{inputenc} % allow utf-8 input
\usepackage[T1]{fontenc}    % use 8-bit T1 fonts
\usepackage{hyperref}       % hyperlinks
\usepackage{url}            % simple URL typesetting
\usepackage{booktabs}       % professional-quality tables
\usepackage{amsfonts}       % blackboard math symbols
\usepackage{nicefrac}       % compact symbols for 1/2, etc.
\usepackage{microtype}      % microtypography
\usepackage{xcolor}         % colors
\usepackage{enumitem} % for customizing lists
\usepackage{amsmath} 
\usepackage{natbib}
\setcitestyle{numbers,square}
\setlist[itemize]{noitemsep, topsep=0pt}
\usepackage{diagbox} 
\usepackage{float}
\usepackage{tablefootnote}
\usepackage{multirow}
\usepackage{multicol}
\usepackage{adjustbox}
\usepackage{subfig}
\usepackage{graphicx}
\usepackage{caption}
\usepackage{longtable}

\usepackage{graphicx}
\usepackage{makecell}
\usepackage{wrapfig}

\title{Stage Light is Sequence$^2$: \\ Multi-Light Control via Imitation Learning}

\author{Zijian Zhao$^1$, Dian Jin$^2$, Zijing Zhou$^3$, Xiaoyu Zhang$^{4}$ \thanks{Corresponding Author: Xiaoyu Zhang (xiaoyu.zhang@cityu.edu.hk)} \\
$^1$The Hong Kong University of Science and Technology $^2$The Hong Kong Polytechnic University \\
$^3$The University of Hong Kong $^4$City University of Hong Kong
}

\begin{document}

\maketitle

\begin{abstract}
Music-inspired Automatic Stage Lighting Control (ASLC) has gained increasing attention in recent years due to the substantial time and financial costs associated with hiring and training professional lighting engineers. However, existing methods suffer from several notable limitations: the low interpretability of rule-based approaches, the restriction to single-primary-light control in music-to-color-space methods, and the limited transferability of music-to-controlling-parameter frameworks. To address these gaps, we propose SeqLight, a hierarchical deep learning framework that maps music to multi-light Hue-Saturation-Value (HSV) space.
% \xz{Need to provide the full name of HSV when first mentioning it}
Our approach first customizes SkipBART, an end-to-end single primary light generation model, 
% \xz{need to briefly introduce SkipBART before mentioning it} 
to predict the full light color distribution for each frame, followed by hybrid Imitation Learning (IL) techniques to derive an effective decomposition strategy that distributes the global color distribution among individual lights. 
% \xz{The following two sentences are too detailed. Please condense and try to combine them into one sentence.} Notably, the light decomposition module can be trained under different venue-specific lighting configurations using only mixed light data, without requiring demonstrations from professional lighting engineers. This design enables our solution to adapt flexibly across diverse venues. 
Notably, the light decomposition module can be trained under varying venue-specific lighting configurations using only mixed light data and no professional demonstrations, thereby flexibly adapting across diverse venues. 
In this stage, we formulate the light decomposition task as a Goal-Conditioned Markov Decision Process (GCMDP), construct an expert demonstration set inspired by Hindsight Experience Replay (HER), and introduce a three-phase IL training pipeline, achieving strong generalization capability. To validate our IL solution for the proposed GCMDP, we conduct quantitative analysis to compare model performance across different training phases, demonstrating that our design effectively improves performance and generalization capacity. Furthermore, we also conduct a human study to evaluate SeqLight by comparing it with competitive baselines on music-conditioned light generation tasks across different music styles. The results show that SeqLight achieves the best overall preference scores in both in-domain and out-of-domain settings. The code and trained parameters of this paper is provided at the anonymous repository \url{https://anonymous.4open.science/r/SeqLight-23EE}.
\end{abstract}

\section{Introduction}

Stage lighting plays a critical role in live music performances, shaping the experience of both performers and audiences. In recent years, Automatic Stage Lighting Control (ASLC) has garnered increasing attention due to its potential not only to reduce reliance on expensive professional lighting engineers but also to inspire amateurs seeking to create expressive stage lighting designs.

Current ASLC methods can be broadly categorized into two main approaches: rule-based solutions and end-to-end solutions. Rule-based methods \cite{mabpa2021automatic,bonde2018auditory,liao2023automatic} typically first segment music pieces into several categories based on attributes such as style, emotion, and chords, and then map each category to predefined light patterns. However, these approaches suffer from limited interpretability \cite{zhao2026automatic,mcdonald2022illuminating} and are constrained by the coarseness and accuracy of the underlying classification models (For instance, the emotion classification task involves only four categories, yet a range of methods achieve accuracies below 80\%, as reported in \cite{zhao2024adversarial}.). To address these limitations, \cite{zhao2026automatic} proposed framing ASLC as an end-to-end art content generation task, learning directly from real-world data produced by lighting engineers. They introduced Skip-BART, trained on their proposed live video dataset Rock, Punk, Metal, and Core - Livehouse Lighting (RPMC-L$^2$), to map music to light hue and value within each frame.\footnote{Note that saturation is fixed at 100\%, as pure colors are predominantly used in practice, which is also adopted in this paper.} Nevertheless, these methods are limited to single primary light control and overlook the variability of lighting configurations across different venues. More recently, \cite{wang2026lightinggen} proposed training a model to directly map music to control parameters (DMX) for each individual light. However, this approach lacks transferability across venues with differing setups, and the cost of collecting professional control data for each venue remains prohibitive. More detailed reviews to related works could be found at Appendix \ref{sec:related work}.

To address the aforementioned challenges, we propose SeqLight, the first color-space multi-light generation solution. Our method decomposes the problem into two sub-tasks: first, we customize Skip-BART to predict the full hue and value distribution of all lights within each frame; second, we train a light distribution decomposition model for each venue using Imitation Learning (IL). This decoupled hierarchical design offers two key advantages: (i) The first stage is independent of venue-specific lighting configurations and relies solely on video data, enabling mixed training across different venues, which is a particularly valuable property for the Music Information Retrieval (MIR) field with limited dataset. (ii) The second stage operates independently of music, eliminating the need for professional lighting engineers in data collection. Specifically, we formulate the light decomposition task as a Goal-Conditioned Markov Decision Process (GCMDP) and introduce a Hindsight Experience Replay (HER)-inspired \cite{andrychowicz2017hindsight} method to collect expert trajectories for IL using only the mixed light data itself. Overall, the proposed method simultaneously addresses the data scarcity and low transferability issues of music-to-controlling-parameter approaches, while achieving multi-light generation, marking a significant advancement in music-to-color-space methods. We evaluate our methodology through both quantitative analysis and human evaluation, and our solution demonstrates consistently promising performance across domains.
% \textbf{\color{orange}The human evaluation results demonstrate that SeqLight achieves the best overall mean scores in both in-domain and out-of-domain settings. Specifically, the proposed SeqLight outperforms the best comparison object 6.9\% and 18.9\% in the two settings, respectively.
% % In the in-domain evaluation, SeqLight obtains an overall score of 4.68, outperforming other baseline methods. In the out-of-domain evaluation, SeqLight also achieves the highest overall score of 4.21.
% }
The main contributions of this paper can be summarized as:
\begin{itemize}[left=0pt]
    \item We propose SeqLight, the first color-space multi-light ASLC method. Specifically, we adapt Skip-BART to predict frame-level hue and value distributions for all lights, and formulate the task of decomposing this global distribution into individual light values and hues as a GCMDP. Additionally, we incorporate each light's previous frame state as a constraint to prevent overshooting and enhance practical control stability.
    
    \vspace{0.1cm}
    \item We introduce a hybrid IL pipeline that that eliminates the need for handcrafted reward functions while learning a powerful policy for the proposed GCMDP.
    % \xz{"To address"? GCMDP is not an issues, so it shouldn't need to be addressed. What's the relationship between the IL pipiline and GCMDP?} 
    Concretely, we first pre-train the policy using Behavioral Cloning (BC), then apply Adversarial Inverse Reinforcement Learning (AIRL) to learn a reward model, and finally fine-tune the policy in more complex scenarios. We also propose a HER-inspired method for generating expert trajectories by deriving goals from real light observations, and introduce a model enhancement technique that predicts hue and value distributions from generated lights as an Auxiliary (AUX) loss. Furthermore, we identify a limitation in conventional AIRL with Actor-Critic (AC) policies, namely the difficulty of training the critic due to the constantly evolving reward model, and address this by incorporating Group Relative Policy Optimization (GRPO), which replaces the critic-based advantage with a group-relative advantage.
    
    \vspace{0.1cm}
    \item We conduct both quantitative analysis and human evaluation to validate our proposed method. The quantitative analysis demonstrates the effectiveness of our proposed three-phase IL framework with GRPO, which improves model performance and generalization capacity. {\color{black}Human evaluation shows that SeqLight achieves the highest overall preference scores in both in-domain and out-of-domain settings. Specifically, our method outperforms the best comparison object 16.4\% and 13.5\% in the two settings, respectively.}

\end{itemize}

\section{Music-Inspired Multi-Light Generation}\label{sec:overview}
% \subsection{Overview} \label{sec:overview}

\begin{figure}[t!]
\centering 
\includegraphics[width=0.8\textwidth]{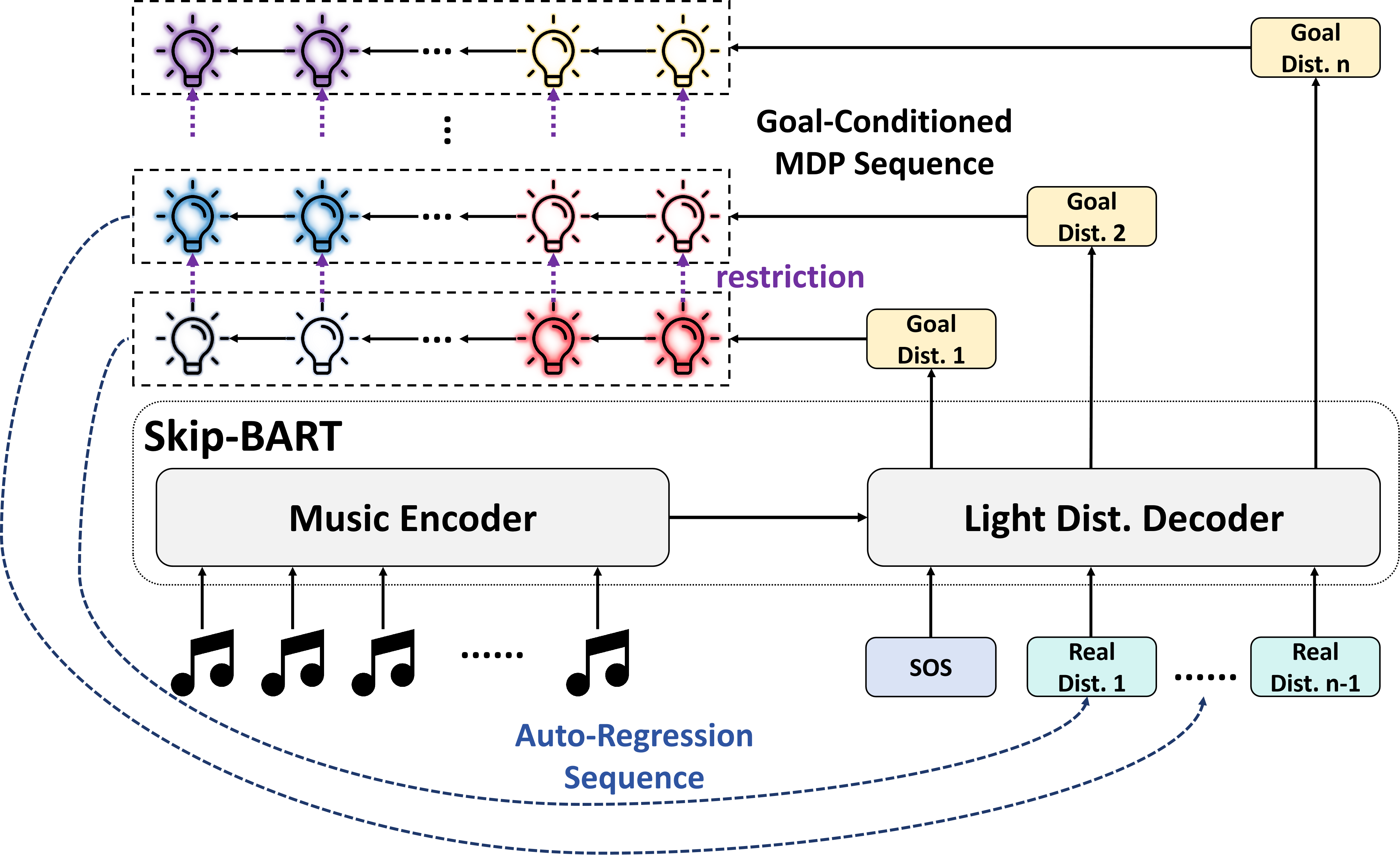}
\caption{Workflow}
% \caption{Workflow: At each step, Skip-BART \cite{zhao2026automatic} generates goal distributions for the hue and value of all lights combined, based on the entire music sequence and the prefix frames' real-generated hue and value distributions. The goal is then decomposed to each light via a GCMDP, where each light's state in the last frame serves as an additional constraint to prevent overshooting.}
\label{fig:workflow}
\end{figure}

The overall workflow of SeqLight is illustrated in Fig. \ref{fig:workflow}. First, the modified Skip-BART \cite{zhao2026automatic} generates the full hue and value distributions for all lights, conditioned on the music and previously generated results. The resulting light distribution is then decomposed into per-light hue and value controls via a GCMDP For simplicity, we assume in this paper that light positions and the control order within each frame are fixed in advance, and that saturation is fixed at 100\%, i.e., we consider only pure colors.

To begin, we emphasize the motivation behind this two-stage hierarchical design. First, to ensure high rationality and interpretability \cite{zhao2026automatic}, we choose to learn from labeled data produced by human lighting engineers rather than relying on rule-based mapping. A naive approach to multi-light control would involve collecting labeled data for each individual light and training an end-to-end model, as in prior music-to-controlling-parameter methods. However, such an approach suffers from low transferability and high data collection costs. Our two-stage solution successfully overcomes these challenges: (i) For the first stage, we observe that in performances involving musicians with professional lighting engineers, the lighting is designed specifically for each piece of music and exhibits similar visual characteristics even across different venues. A real-world example is provided at Appendix \ref{sec:mekader}.

We now formally define the music-inspired multi-light generation task. For a given music piece consisting of $\mathcal{T}$ frames $X=\{ x_1, x_2, \ldots, x_{\mathcal{T}} \}$, our goal is to find a light control sequence $\mathcal{U}=\{ U_1, U_2, \ldots, U_{\mathcal{T}} \}$ that best approximates the real full-light distributions $Y=\{ y_1, y_2, \ldots, y_{\mathcal{T}} \}$. At the $j$-th frame, $U_j$ is defined as $U_j=[u_{1,j}, u_{2,j}, \ldots, u_{n,j}]$, where $n$ is the number of lights. For the $i$-th light, the control action is $u_{i,j}=[h_{i,j}, v_{i,j}]$, with hue $h_{i,j}\in[0,2\pi)$ and value $v_{i,j}\in[0,1]$. The ground-truth for frame $j$, denoted $y_j$, consists of a hue distribution $H_j\in[0,1]^{b_h}$ and a value distribution $V_j\in[0,1]^{b_v}$, where hue and value are discretized into $b_h$ and $b_v$ bins, respectively, with $\sum_{i=1}^{b_h} H_{j,i} = \sum_{i=1}^{b_v} V_{j,i} = 1$. Our optimization objective is then defined as:
\begin{equation}
\begin{aligned}
& \min_{\Theta}  \mathbb{E}_{\mathcal{U}}\big[ \mathrm{Dist}\big(\mathrm{Mix}(\mathcal{U}) || y_j\big) \big],\\
\text{s.t.}\quad & U_j = \mathrm{F}\big(X, U_{:j-1}; \Theta\big),
\end{aligned}
\label{eq:objective}
\end{equation}
where $\mathrm{F}(\cdot,\cdot;\Theta)$ is the mapping function parameterized by $\Theta$, $\mathrm{Mix}(\cdot)$ denotes the mixed (aggregated) hue and value distribution induced by the per-light controls $U_j$, and $\mathrm{Dist}(\cdot||\cdot)$ is a chosen distributional distance (e.g., L1, Wasserstein, KL). Further details are provided in Appendix \ref{sec:Preliminary}.

Although it may appear possible to directly train a network by deriving a loss function from Eq. \eqref{eq:objective} and applying gradient descent, this approach is infeasible in practice. First, the solution is not unique: many different combinations of per-light controls can yield the same aggregated distribution, making the learning problem more challenging than standard supervised regression. Second, even if the light-mixing process could be simulated, it often involves complex and potentially non-differentiable operations, which hinder end-to-end gradient-based optimization. These challenges further underscore the necessity of our proposed hierarchical solution.

\subsection{Training Process}

In the first stage, we adapt the original Skip-BART to predict the light distributions for each frame. We employ KL divergence as the distance metric and define the supervised loss as
\begin{equation}
\begin{aligned}
L_{\mathrm{sup}}^{\phi} &= \mathbb{E}_{B^t}\big[ \mathrm{KL}\big(\hat{H}_j \,\|\, H_j\big) +\mathrm{KL}\big(\hat{V}_j \,\|\, V_j\big)  \big],\\
\hat{H}_j, \hat{V}_j &= \mathrm{Skip\text{-}BART}\big(X,\;[\hat{H}_{:j-1},\,\hat{V}_{:j-1}];\phi\big),
\end{aligned}
\label{eq:skip-bart}
\end{equation}
where $\phi$ denotes the network parameters, $B^t$ represents training set, and the direction of the KL divergence is chosen following the principle of Variational Autoencoders (VAE) \cite{kingma2013auto}. Concretely, we modify Skip-BART by replacing its input and output projection layers with new MLPs to match the dimensionality of the full light distributions (as opposed to the single primary light predicted by the original model). For the backbone, we initialize $\phi$ using the pretrained Skip-BART weights from \cite{zhao2026automatic}, which reduces training time and improves convergence via transfer learning \cite{pan2009survey}.

In the second stage, we decompose the predicted hue and value distributions into per-light controls within each frame by formulating the task as a GCMDP and solving it with Reinforcement Learning (RL). The objective is to maximize the expected cumulative reward:
\begin{equation}
\label{eq:rl_goal}
\max_{\theta} \mathscr{J}(\theta)= \max_{\theta} \mathbb{E}_{\pi_\theta}\Big[ \sum_{t=1}^n \gamma^{t-1} \mathscr{R}(s_t,a_t,g) \Big],
\end{equation}
where $\mathscr{R}(\cdot,\cdot,\cdot)$ is the reward function, $\pi_\theta$ is the policy parameterized by $\theta$, $s_t$ and $a_t$ are the state and action at time $t$, $g$ denotes the goal, and $\gamma$ is the discount factor. In our task, it is straightforward to evaluate whether a complete trajectory is good by comparing the final aggregated distribution to the goal, but it is difficult to design informative stepwise reward signals. This reward sparsity hinders efficient RL training. To address this, we adopt a hybrid IL approach: we first generate expert trajectories using HER, then pre-train the policy with BC, next employ AIRL \cite{fu2018learning} to learn a reward model from expert and policy trajectories, and finally fine-tune the policy using the learned reward to enable generalization beyond the expert demonstrations. Further details of our goal-conditioned light decomposition method are provided in Section \ref{sec:goal}.

\subsection{Framework Pipeline} \label{sec:pipeline}
After obtaining the well-trained Skip-BART for full-light distribution generation and the light-decomposition policy, we use SeqLight to generate lights as illustrated in Fig. \ref{fig:workflow}. At each frame $j$, Skip-BART first predicts the hue and value distributions conditioned on the music and the previously generated frames:
\begin{equation}
\begin{aligned}
\hat{H}_j, \hat{V}_j &= \mathrm{Skip\text{-}BART}\big(X,\;[\tilde{H}_{:j-1},\,\tilde{V}_{:j-1}];\phi^*\big),\\
\tilde{H}_{k},\,\tilde{V}_{k} &= \mathrm{Mix}(U_k),
\end{aligned}
\label{eq:skip-bart-generate}
\end{equation}
where $\tilde{H}_{k},\tilde{V}_{k}$ denote the aggregated hue and value distributions in frame $k$ resulting from the mixing function $\mathrm{Mix}(\cdot)$.

A naive approach to obtain $U_j$ is to sample directly from $\pi_{\theta^*}$ conditioned on the goal $g_j=[\hat{H}_j,\hat{V}_j]$. However, this approach has two drawbacks: (i) it may lead to overshooting of individual lights' previous states; and (ii) it can produce temporally inconsistent control sequences across frames, since many different per-light decompositions can realize the same goal distribution. To mitigate these issues, we adopt a last-frame state constrained sampling strategy:
\begin{equation}
\begin{aligned}
h_{i,j}, v_{i,j} &\sim \pi_{\theta^*}(s_{i,j}, g_j, \iota),\\
\text{s.t.}\quad & \mathrm{D}_h\big(h_{i,j} \,\|\, h_{i,j-1}\big) < d^h,\\
& \mathrm{D}_v\big(v_{i,j} \,\|\, v_{i,j-1}\big) < d^v,
\end{aligned}
\label{eq:restriction-sample}
\end{equation}
where $s_{i,j}$ is the state for light $i$ at step $j$, $\iota$ is a sampling temperature parameter that controls diversity, $d^h$ and $d^v$ are the maximum allowable changes for hue and value to prevent overshooting and ensure temporal smoothness, and $\mathrm{D}_h,\mathrm{D}_v$ are distance metrics (we adopt the same metrics used in \cite{zhao2026automatic}, detailed in Appendix \ref{sec:Preliminary}).

\section{Goal-Conditioned Light Decomposition}  \label{sec:goal}
\subsection{Problem Setup}

\begin{wrapfigure}{r}{0.6\textwidth}
% \begin{figure}[htbp]
\centering 
\includegraphics[width=0.6\textwidth]{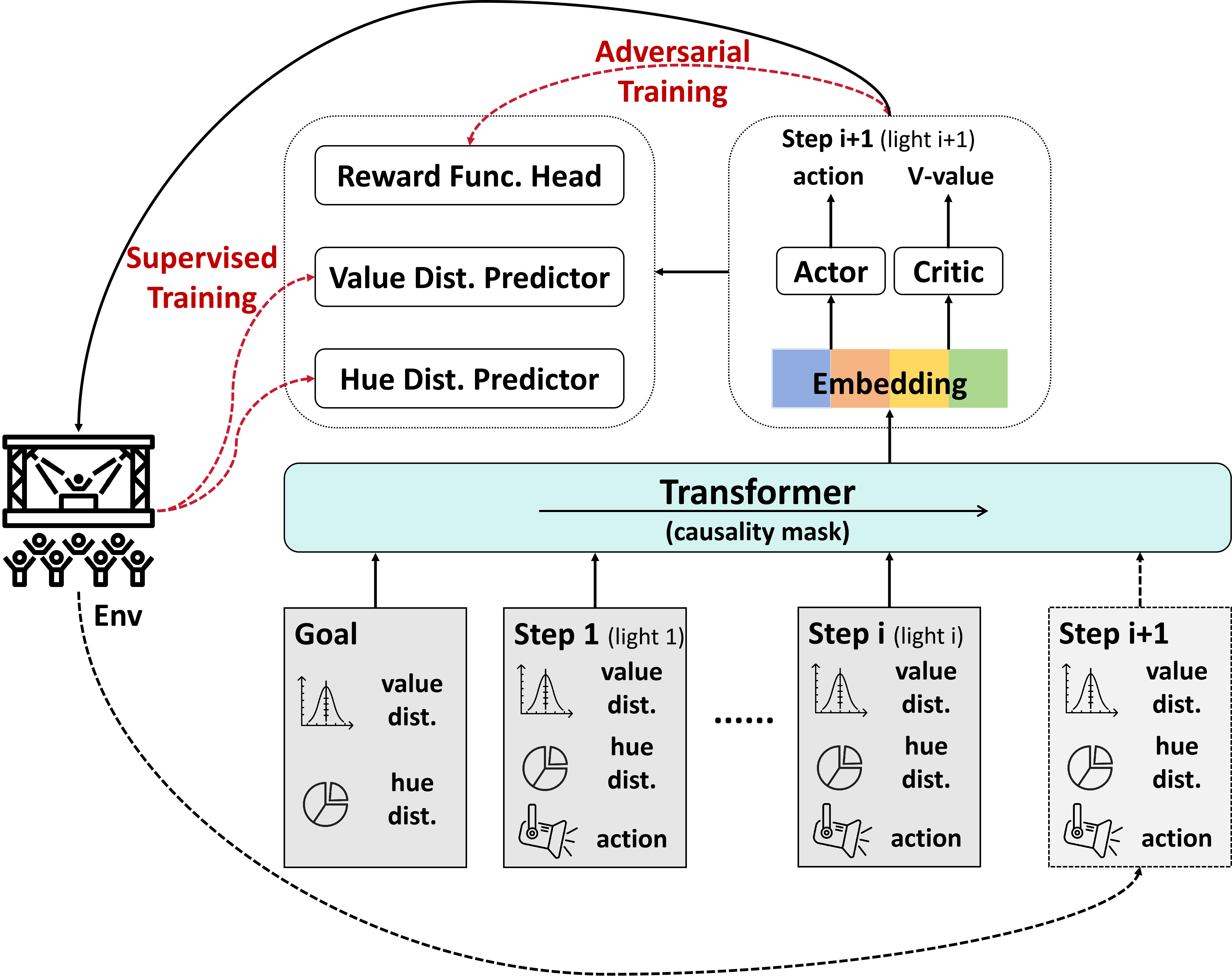}
% \caption{Network Architecture: Our network takes the goal and historical trajectory as input, utilizing a GPT-style Transformer \cite{vaswani2017attention} for feature extraction. The output embedding is forwarded to both the actor and critic for action generation and value function estimation. This embedding, combined with the action, is then fed into a reward model for reward estimation, as well as to value distribution and hue distribution heads for predicting the new distribution after taking the new action, as AUC.}
\caption{Network Architecture}
\label{fig:main}
% \end{figure}
\end{wrapfigure}

We formulate the light-decomposition task as a GCMDP, defined by the tuple $\langle S, A, R, P, \gamma, G, \rho_g \rangle$, where $S$, $A$, $R$, $P$, and $\gamma$ denote the state space, action space, reward function, transition function, and discount factor, respectively, as in a conventional MDP, while $G$ and $\rho_g$ represent the goal space and the goal distribution.

\textbf{(i) State:} At each step $t$, the state $s_t$ is defined as $s_t = [s_{:t-1}, a_{t-1}, \mathrm{Mix}(a_{:t-1})]$, where $s_{:t-1}$ and $a_{t-1}$ comprise the history of states and actions up to step $t-1$, and $\mathrm{Mix}(a_{:t-1})$ denotes the aggregated hue and value distribution generated by the first $t-1$ actions.

\textbf{(ii) Action:} he action at step $t$, denoted $a_t$, corresponds to the per-light control $U$ introduced in the previous section and consists of hue and value components: $a_t = [a_t^h, a_t^v]$. In this work, we adopt a stochastic policy and sample actions according to $a_t \sim \pi_\theta(s_t, g)$. Since hue is circular with domain $[0,2\pi)$, we model it using a Von Mises distribution. For value, which lies in $[0,1]$, we employ a Beta distribution. The detailed formal definition is provided in Appendix \ref{sec:Preliminary}.

\textbf{(iii) Reward function:} As discussed previously, designing a stepwise reward for this task is challenging. Therefore, we parameterize the reward as a neural network $\mathscr{R}_{\Phi}(s_t,a_t,g): S \times A \times G \rightarrow \mathbb{R}$ with parameters $\Phi$. The reward model is learned via AIRL, which will be described in detail later.

\textbf{(iv) State transition function:} Although we adopt a model-free RL approach in this paper, the environment transition $\mathscr{P}(\cdot \mid s_t, a_t)$ is actually deterministic and known: given an action, the resulting aggregated hue and value distribution can be computed, and the next state is formed accordingly. We exploit this physical regularity by incorporating it as an AUX loss to enhance feature learning in both the policy and reward networks.

\textbf{(v) Goal and goal distribution:} The goal is defined as $G=[H,V]$, i.e., the target full-light hue and value distribution that the aggregated per-light controls are intended to match. We consider two types of goal distributions: $\rho_g^e$, corresponding to expert trajectories, and $\rho_g^a$, corresponding to arbitrary goals. Further details are provided in the following sections.

In this paper, we use $\mathscr{V}$, $\mathscr{Q}$, and $\mathscr{A}$ to denote the state-value function, action-value function, and advantage function, respectively, for standard goal-conditioned RL. Their formal definitions are provided in Appendix \ref{sec:Preliminary}.

% \subsection{Neural Network Architecture}

To solve the proposed GCMDP, we adopt a Transformer-based network \cite{vaswani2017attention}, as illustrated in Fig. \ref{fig:main}. Below, we summarize the network architecture, with further details provided in Appendix \ref{sec:network}.
\begin{equation}
\begin{aligned}
a_t \sim \pi_{\theta}(s_t, g), \quad
\Upsilon_t = \mathscr{V}_{\psi}(s_t, g),  \quad
r_t = \mathscr{R}_{\Phi}(s_t, a_t, g),  \quad
(\mathcal{H}_{t}, \mathcal{V}_{t}) = \mathrm{P}_{\Psi}(s_t, a_t),
\end{aligned}
\label{eq:network}
\end{equation}
where $\Upsilon_t$ denotes the estimated state value, $\mathcal{H}_{t}$ and $\mathcal{V}_{t}$ denote the predicted hue and value distributions after executing the action, and $\theta,\psi,\Phi,\Psi$ are associated with different network heads but share the same Transformer backbone. Note that although all outputs implicitly depend on $(s, a, g)$, we retain only the relevant arguments in each function definition to avoid notational clutter.

\subsection{HER-Inspired Expert Trajectory Generation}

In our method, expert trajectories are generated following the HER principle \cite{andrychowicz2017hindsight}. We first sample random hue and value actions for each light to obtain a trajectory $\langle s_1, a_1, s_2, a_2, \ldots, s_n, a_n \rangle$. The goal for this expert trajectory is then labeled as the aggregated result of all lights, i.e., the optimal goal $g = \mathrm{Mix}(a_{:n})$. During training, we also generate additional expert trajectories by relabeling the goals of trajectories sampled from the current policy. Specifically, we denote the expert dataset as $B^{e}=\{\tau^{e}_{1},\tau^{e}_{2},\ldots,\tau^{e}_{N}\}$, where $\tau^{e}_{i}$ is a trajectory annotated with its goal and $N$ is the dataset size.

The goals of expert trajectories therefore originate from real mixed lights and follow the distribution $\rho_{g}^e$. In contrast, during the light control phase (Section \ref{sec:pipeline}), the goal is provided by the Skip-BART predictor, which is trained on noisy real-world data. Moreover, prediction errors introduce bias, causing these goals to deviate from $\rho_{g}^e$. To account for this discrepancy, we introduce an alternative goal distribution $\rho_{g}^a$, in which hue and value are sampled completely at random rather than computed from real lights. After training the policy via IL on expert goals sampled from $\rho_{g}^e$, we fine-tune it on arbitrary goals drawn from $\rho_{g}^a$ to enhance robustness and generalization capacity.

\subsection{Training Process}
\subsubsection{Phase 1: Policy Pre-Training by BC}

To initialize the policy $\pi_{\theta}$ with expert behavior, we pre-train it via BC by maximizing the likelihood of expert actions (equivalently, minimizing the negative log-likelihood). The BC loss is defined as:
\begin{equation}
\begin{aligned}
L_{\mathrm{bc}}^\theta &= - \mathbb{E}_{B^e}\big[ \log \pi_{\theta}(a \mid s,g) \big].
\end{aligned}
\label{eq:bc}
\end{equation}

Additionally, we incorporate an AUX loss from the hue and value predictor to enhance the model's feature extraction and transition modeling capabilities. This AUX loss follows the same supervised formulation used in Skip-BART (Eq. \eqref{eq:skip-bart}):
\begin{equation}
\begin{aligned}
L_{\mathrm{aux}}^{\Psi} &= \mathbb{E}_{B^e}\Big[ \mathrm{KL}\big(\mathcal{H}_{t} \,\|\, \mathrm{Mix}(a_{1:t}).\text{hue} \big) +\mathrm{KL}\big(\mathcal{V}_{t} \,\|\, \mathrm{Mix}(a_{1:t}).\text{value} \big) \Big],
% \\ \mathcal{H}_{t}, \mathcal{V}_{t} &= \mathrm{P}_{\Psi}(s_t, a_t).
\end{aligned}
\label{eq:aux}
\end{equation}
where $\mathcal{H}_{t}, \mathcal{V}_{t}$ come from Eq. \ref{eq:network}. In summary, the combined loss for Phase 1 is defined as:
\begin{equation}
\begin{aligned}
\mathcal{L}_1^{\theta,\Psi} = L_{\mathrm{bc}}^\theta + \eta\, L_{\mathrm{aux}}^{\Psi},
\end{aligned}
\label{eq:phase1}
\end{equation}
where $\eta \ge 0$ is a weighting coefficient.

\subsubsection{Phase 2: Reward Model Training by AIRL}

In AIRL, we alternately train a discriminator and a policy. Following the Generative Adversarial Network - Guided Cost Learning (GAN-GCL) framework \cite{finn2016connection}, the discriminator is defined as:
\begin{equation}
\mathscr{D}(\tau) = \frac{\exp(\mathscr{F}(\tau))}{\exp(\mathscr{F}(\tau)) + \pi(\tau)},
\label{eq:discriminator}
\end{equation}
where $\mathscr{D}(\tau)$ denotes the probability that trajectory $\tau$ originates from the expert, and $\mathscr{F}(\cdot)$ is a learned scoring function (note that $\tau$ may be of arbitrary length). The reward induced by the discriminator is given by:
\begin{equation}
\mathscr{R}(\tau) = \log \mathscr{D}(\tau) - \log\big(1-\mathscr{D}(\tau)\big) = \mathscr{F}(\tau) -\log \pi(\tau),
\label{eq:reward}
\end{equation}
which can be rearranged into the equivalent form:
\begin{equation}
\mathscr{D}(\tau) = \frac{1}{1+\exp(-\mathscr{R}(\tau))} = \operatorname{Sigmoid}\big(\mathscr{R}(\tau)\big).
\label{eq:discriminator2}
\end{equation}
Since the reward function is parameterized by $\Phi$, we write the discriminator as $\mathscr{D}_{\Phi}(\cdot)$. The discriminator loss follows the standard GAN objective \cite{goodfellow2014generative}:
\begin{equation}
L_{\mathrm{dis}}^{\Phi} = -\mathbb{E}_{B^e}\big[\log \mathscr{D}_{\Phi}(s,a,g )\big]
- \mathbb{E}_{\pi_{\theta}}\big[\log\big(1-\mathscr{D}_{\Phi}(s,a,g )\big)\big],
\label{eq:discriminator_loss}
\end{equation}
where we use $(s,a,g)$ rather than $(s,a,s',g)$ because our state transitions are deterministic, and $s'$ denotes the next state.

The policy is trained alternately with the discriminator. For our stochastic continuous control problem, we adopt PPO \cite{schulman2017proximal} as a common choice. The PPO actor and critic losses are defined using the policy-gradient objective and temporal-difference error, respectively:
\begin{equation}
\begin{aligned}
L_{\mathrm{actor\text{-}PPO}}^{\theta} &= \mathbb{E}_{\pi_{\theta^-}}\Big[ \min\Big( \frac{\pi_{\theta}(s,a)}{\pi_{\theta^-}(s,a)} \,\tilde{\mathscr{A}}_{\psi,\Phi}(s,a,g),\; 
\operatorname{CLIP}\big(\frac{\pi_{\theta}(s,a)}{\pi_{\theta^-}(s,a)} ,1-\epsilon,1+\epsilon\big)\,\tilde{\mathscr{A}}_{\psi,\Phi}(s,a,g) \Big) \Big], \\
L_{\mathrm{critic\text{-}PPO}}^{\psi} &= \mathbb{E}_{\pi_{\theta^-}}\Big[ \big(\mathscr{R}_{\Phi}(s,a,g) + \gamma \mathscr{V}_{\psi}(s',g) - \mathscr{V}_{\psi}(s,g)\big)^{2} \Big],
\end{aligned}
\label{eq:ppo}
\end{equation}
where $\pi_{\theta^-}$ is the behavior policy used for data collection, $\epsilon$ is the PPO clipping hyper-parameter, and $\tilde{\mathscr{A}}_{\psi,\Phi}$ denotes the Generalized Advantage Estimation (GAE) \cite{schulman2015high} computed using the value function $\mathscr{V}_{\psi}$ and the reward model $\mathscr{R}_{\Phi}$. The overall loss for the PPO-based Phase 2 is therefore:
\begin{equation}
\mathcal{L}_{\mathrm{2\text{-}PPO}}^{\Phi,\theta,\psi,\Psi} = L_{\mathrm{dis}}^{\Phi} + \mathbb{E}_{g \sim \rho_g^e} \big[ L_{\mathrm{actor\text{-}PPO}}^{\theta} + L_{\mathrm{critic\text{-}PPO}}^{\psi} \big] + \delta\, L_{\mathrm{bc}}^{\theta} + \eta\, L_{\mathrm{aux}}^{\Psi},
\label{eq:phase2-ppo}
\end{equation}
where $\delta,\eta \ge 0$ are weighting coefficients, and $L_{\mathrm{bc}}^{\theta}$ and $L_{\mathrm{aux}}^{\Psi}$ serve as auxiliary losses in this phase. (Note that, except for the policy loss in Phase 3, all other components use goals drawn exclusively from the expert distribution $\rho_{g}^{e}$, which is already incorporated into the respective loss terms.)

However, in the PPO-based solution, we observe strong interactions among the reward model, critic, and actor during this phase. If the critic fails to keep pace with changes in the reward model (whose scale may drift during training), actor updates can become misdirected and unstable, which in turn degrades the reward model (since the policy also influences the reward defined in Eq. \eqref{eq:reward}). To mitigate this issue, we adopt Group Relative Policy Optimization (GRPO) \cite{shao2024deepseekmath} as an alternative. For each goal, GRPO samples multiple trajectories and replaces the advantage with a group-relative reward-to-go. The GRPO-style advantage function, denoted as $\overline{\mathscr{A}}$, is formally defined in Appendix \ref{sec:Preliminary}. Then the GRPO actor loss replaces the PPO advantage with this group-relative advantage:
\begin{equation}
L_{\mathrm{actor\text{-}GRPO}}^{\theta} = \mathbb{E}_{\pi_{\theta^-}}\Big[ \min\Big( \frac{\pi_{\theta}(s,a)}{\pi_{\theta^-}(s,a)} \,\overline{\mathscr{A}}_{\Phi}(s,a,g),\;
\operatorname{CLIP}\big(\frac{\pi_{\theta}(s,a)}{\pi_{\theta^-}(s,a)}),1-\epsilon,1+\epsilon\big)\,\overline{\mathscr{A}}_{\Phi}(s,a,g) \Big) \Big].
\label{eq:grpo}
\end{equation}
By eliminating the critic and using group-relative returns, GRPO reduces sensitivity to reward scaling and focuses updates on trajectories that perform better within the sampled group. The overall loss for the GRPO-based Phase 2 is therefore:
\begin{equation}
\mathcal{L}_{\mathrm{2\text{-}GRPO}}^{\Phi,\theta,\Psi} = L_{\mathrm{dis}}^{\Phi} + \mathbb{E}_{g \sim \rho_g^e} \big[L_{\mathrm{actor\text{-}GRPO}}^{\theta}] + \delta\, L_{\mathrm{bc}}^{\theta} + \eta\, L_{\mathrm{aux}}^{\Psi}.
\label{eq:phase2-grpo}
\end{equation}

\subsubsection{Phase 3: Policy Fine-Tuning by RL}

In Phase 3, we aim to further improve the policy's generalization capacity, particularly for goals outside the expert trajectory distribution. To this end, we freeze the reward model and train only the policy network. Since the policy and reward model share the same Transformer backbone, we must fix the backbone parameters to preserve the reward model unchanged. Let $\theta'$ and $\psi'$ denote the parameters of the policy and critic, respectively, that are external to the shared Transformer backbone. The loss functions for this phase are therefore:
\begin{equation}
\begin{aligned}
\mathcal{L}_{\mathrm{3\text{-}PPO}}^{\theta',\psi'} &= \mathbb{E}_{g \sim \rho_g^a} \big[L_{\mathrm{actor\text{-}PPO}}^{\theta'} + L_{\mathrm{critic\text{-}PPO}}^{\psi'}] + \delta\, L_{\mathrm{bc}}^{\theta'}, \\
\mathcal{L}_{\mathrm{3\text{-}GRPO}}^{\theta'} &=  \mathbb{E}_{g \sim \rho_g^a} \big[L_{\mathrm{actor\text{-}GRPO}}^{\theta'}] + \delta\, L_{\mathrm{bc}}^{\theta'}.
\end{aligned}
\label{eq:phase3}
\end{equation}
Note that since the backbone parameters are fixed in this phase, adding the AUX loss is neither necessary nor effective.

\subsection{Post-Processing}

In the light decomposition task, determining the value for each light requires considering both how to achieve the target hue distribution and how scaling a light's value affects its contribution to the overall mixture. Fortunately, scaling the values of all lights by a common factor preserves the relative hue distribution of the full light set. Leveraging this property, we propose a post-processing method that introduces a scaling factor $f$ to adjust the generated light values, aiming to make the resulting mixture's value distribution as close as possible to the goal value distribution. This can be formulated as the following optimization problem:
\begin{equation}
\begin{aligned}
& \min_{f \geq 0}  \operatorname{Dist}\Big( \operatorname{Mix}\big( [a_1^h, a_1^v f ], [a_2^h, a_2^v f ], \ldots, [a_n^h, a_n^v f ] \big).\text{value} \,\|\, g^v \Big), \\
\text{s.t.} \quad & 0 \leq a_i^v f \leq 1 \quad \forall i = 1, 2, \ldots, n,
\end{aligned}
\label{eq:scale}
\end{equation}
where $g = [g^h, g^v]$ denotes the target hue and value distributions.

\section{Experiment}

To validate our proposed method, in this section we conduct both quantitative experiments and a human study, comparing our approach with several baselines and ablation variants. Specifically, we train the Skip-BART model for full light distribution prediction using the PMRC-L$^2$ dataset \cite{zhao2026automatic}. To avoid information leakage, we adopt the same training and testing set division as the original Skip-BART \cite{zhao2026automatic}. For the light decomposition component, we utilize a simulation environment equipped with eight circular point lights. Further details regarding the experimental setup, including the simulation configuration, dataset information, and model parameters, are provided in Appendix \ref{sec:experiment}.

\subsection{Quantitative Analysis}

In this section, we first illustrate the efficiency of our goal-conditioned light decomposition method (Section \ref{sec:goal}). We consider two different types of goals: the expert-based goal (In-Domain, ID), which involves sampling the hue and value distribution from a mixture of real lights, and the randomly generated goal (Out-Of-Domain, OOD), which involves directly generating random distributions for both hue and value. We compare the model performance under different phases in training using PPO \cite{schulman2017proximal} and GRPO \cite{shao2024deepseekmath} as policy strategies, respectively. The evaluation metrics include L1 distance, Wasserstein distance, JS divergence, KL divergence, Bhattacharyya distance, and cosine similarity. Based on the results in Table \ref{tab:quant}, we observe that our proposed Phase 3 (GRPO) achieves the best performance in the OOD scenario, while Phase 2 (GRPO) performs best in the ID scenario. This suggests that although Phase 3 enhances generalization capacity, it also sacrifices some degree of overfitting capability. A more detailed experimental analysis along with ablation studies can be found in Appendix \ref{sec:Quantitative Analysis}.

\begin{table*}[htbp]
\centering
\caption{Model Performance on Goal-Conditioned Light Decomposition Task: The best results are highlighted in \textbf{bold}, while the second-best results are indicated with \underline{underlining}. This notation is consistent throughout the following tables. (Mean ± Standard Deviation (M ± SD))}
\begin{adjustbox}{width=0.9\textwidth}
\begin{tabular}{lcccccc}
\toprule
\multirow{2}{*}{\textbf{Model}} & \multicolumn{2}{c}{\textbf{L1 ($\times 10^{-3}$) $\downarrow$}} & \multicolumn{2}{c}{\textbf{Wasserstein ($\times 10^{-2}$) $\downarrow$}} & \multicolumn{2}{c}{\textbf{JS ($\times 10^{-1}$) $\downarrow$}} \\
\cmidrule(lr){2-3} \cmidrule(lr){4-5} \cmidrule(lr){6-7}
 & \textbf{ID} & \textbf{OOD} & \textbf{ID} & \textbf{OOD} & \textbf{ID} & \textbf{OOD} \\
\midrule
\multicolumn{7}{c}{\textbf{Hue}} \\
\midrule
Phase 1 & 3.58±1.01 & 3.20±0.76 & 5.61±5.70 & 7.99±0.60 & 3.18±1.25 & 2.80±0.93 \\
Phase 2 (GRPO) & \underline{2.66±0.80} & 2.99±0.59 & 4.97±4.96 & \underline{6.89±0.50} & \underline{2.17±0.93} & 2.71±0.93 \\
Phase 3 (GRPO) & 2.73±0.09 & \textbf{2.59±0.89} & 5.19±5.58 & 7.54±5.75 & 2.23±1.08 & \textbf{2.19±1.11} \\
Phase 2 (PPO) & \textbf{2.52±0.87} & \underline{2.70±0.58} & \textbf{4.73±5.51} & \textbf{6.93±5.22} & \textbf{2.03±1.00} & \underline{2.39±0.19} \\
Phase 3 (PPO) & 2.74±0.84 & 3.18±0.53 & \underline{4.80±4.57} & 7.16±4.34 & 2.20±1.02 & 2.87±0.72 \\
\midrule 
\multicolumn{7}{c}{\textbf{Value}} \\
\midrule
Phase 1 & 10.21±3.06 & 11.25±1.46 & 5.98±3.38 & 6.87±0.83 & 1.67±0.67 & 2.44±0.43 \\
Phase 2 (GRPO) & \underline{8.63±3.04} & \underline{9.40±1.82} & \underline{5.05±3.32} & \textbf{5.76±1.30} & \underline{1.31±0.65} & \underline{1.87±0.48} \\
Phase 3 (GRPO) & 9.24±3.32 & \textbf{9.14±2.16} & 5.32±3.44 & \underline{5.78±0.79} & 1.50±0.71 & \textbf{1.85±0.53} \\
Phase 2 (PPO) & \textbf{8.07±2.62} & 10.73±2.20 & \textbf{4.74±3.12} & 6.33±1.36 & \textbf{1.21±0.55} & 2.31±6.55 \\
Phase 3 (PPO) & 9.70±3.26 & 11.50±2.53 & 5.60±3.45 & 7.08±1.13 & 1.52±0.70 & 2.57±9.29 \\
\midrule \midrule
\multirow{2}{*}{\textbf{Model}} & \multicolumn{2}{c}{\textbf{KL $\downarrow$}} & \multicolumn{2}{c}{\textbf{Bhattacharyya ($\times 10^{-1}$) $\downarrow$}} & \multicolumn{2}{c}{\textbf{Cosine ($\times 10^{-1}$) $\uparrow$}} \\
\cmidrule(lr){2-3} \cmidrule(lr){4-5} \cmidrule(lr){6-7}
 & \textbf{ID} & \textbf{OOD} & \textbf{ID} & \textbf{OOD} & \textbf{ID} & \textbf{OOD} \\
\midrule
\multicolumn{7}{c}{\textbf{Hue}} \\
\midrule
Phase 1 & 1.66±1.56 & \underline{1.15±0.52} & 5.79±2.88 & 4.74±2.00 & 6.64±1.45 & 6.61±1.33 \\
Phase 2 (GRPO) & 3.78±4.64 & 1.98±0.89 & \underline{3.49±1.72} & 4.45±2.13 & 7.13±1.29 & 6.23±1.36 \\
Phase 3 (GRPO) & \textbf{1.65±2.33} & \textbf{1.05±0.66} & 3.67±2.06 & \textbf{3.61±2.17} & \underline{7.53±1.24} & \textbf{7.19±1.40} \\
Phase 2 (PPO) & \underline{2.58±3.46} & 1.51±5.00 & 3.19±1.87 & \underline{3.87±1.74} & \textbf{7.56±1.21} & \underline{7.18±1.19} \\
Phase 3 (PPO) & 3.77±4.93 & 1.93±6.65 & \textbf{3.05±2.09} & 4.69±1.64 & 7.11±1.27 & 6.42±1.64 \\
\midrule 
\multicolumn{7}{c}{\textbf{Value}} \\
\midrule
Phase 1 & 0.58±0.24 & 0.86±0.17 & 2.46±1.14 & 3.78±0.88 & 8.00±0.91 & 6.97±0.60 \\
Phase 2 (GRPO) & \underline{0.45±0.23} & \textbf{0.65±0.19} & \underline{1.85±1.07} & \underline{2.78±0.88} & \textbf{8.27±0.86} & \textbf{7.66±0.66} \\
Phase 3 (GRPO) & 0.52±0.25 & \underline{0.66±0.21} & 2.17±1.20 & \textbf{2.70±0.98} & 8.09±0.89 & \underline{7.45±0.74} \\
Phase 2 (PPO) & \textbf{0.44±0.20} & 0.82±0.26 & \textbf{1.64±1.09} & 3.60±1.26 & \underline{8.15±0.75} & 7.13±0.85 \\
Phase 3 (PPO) & 0.52±0.25 & 0.95±0.45 & 2.23±1.17 & 4.28±2.45 & \underline{8.15±0.91} & 6.79±1.24 \\
\bottomrule
\end{tabular}
\end{adjustbox}
\label{tab:quant}
\end{table*}

\subsection{Human Evaluation}

To better assess the alignment of our results with human preferences, we conducted a human evaluation study. Specifically, we asked 30 participants to rate six music pieces paired with lighting sequences generated under different styles. Each piece was evaluated across six metrics \citep{erdmann2025development}, with scores ranging from 1 to 7 (higher scores indicating better quality). The lighting conditions were generated from four sources: Ground Truth, SeqLight, Skip-BART \cite{zhao2026automatic}, and a rule-based method. To evaluate the generalization capability of our approach, we selected three music pieces from the test set of RPMC-L$^2$ \cite{zhao2026automatic} (ID group), covering rock, metal, and core genres. Additionally, we included three pieces from the OOD experiment of Skip-BART, which were generated by Suno \cite{suno} (OOD group). Note that the OOD group does not have corresponding ground-truth lighting. The results of the human evaluation are presented in Table \ref{tab:my_table}, demonstrating that our proposed method consistently achieves promising performance across different music styles and evaluation dimensions. {\color{black}On the ID group, SeqLight achieves the highest overall score ($4.54 \pm 0.88$), outperforming other baseline methods: Skip-BART ($3.90 \pm 0.84$), rule-based ($2.70 \pm 1.26$). It obtains the best mean score on almost all evaluation dimensions. Pairwise comparisons (Table \ref{tab:merged_stats}) further show that SeqLight significantly outperforms Skip-BART in Impact, Rhythm, Surprise, and Overall, and consistently surpasses the rule-based method across all metrics. On the OOD group, SeqLight also achieves the highest scores across all metrics, with an overall score of $3.94 \pm 1.32$ compared with $3.47 \pm 1.01$ for Skip-BART and $2.70 \pm 1.36$ for the rule-based method. These results indicate that SeqLight captures music-lighting correspondence better and maintains stronger generalization ability across unseen music styles.} More details regarding the human evaluation setup and complete results can be found in Appendix \ref{sec:Human Evaluation}.

\section{Conclusion}
In this paper, we propose SeqLight, the first music-to-color-space multi-light ASLC method. Our approach addresses the challenges of low transferability and dataset scarcity through a hierarchical design. First, we train Skip-BART on mixed-venue live videos to predict the full HV distribution of the lights. Subsequently, we employ IL to derive an effective decomposition strategy that maps the predicted distribution to individual light controls. This decomposition task is formulated as a GCMDP and trained independently for each venue, with expert data collected via simple light mixing and goal labeling using HER.
% \textbf{\color{orange}Quantitative experiments demonstrate the effectiveness of SeqLight in generating expressive multi-color lighting distributions conditioned on music, and decomposing the predicted global color distribution into feasible individual-light controls. Human evaluations further show that SeqLight achieves the highest preference scores in both in-domain and out-of-domain settings, while remaining comparable to human-designed ground truth.}
Both quantitative analysis and human evaluation demonstrate the generalization and efficiency of the proposed method across both in-domain and out-of-domain settings. Further discussions can be found in Appendix~\ref{sec:Discussions}.

\begin{table*}[htbp]
    \centering
    \caption{Human Evaluation Scores}
    \label{tab:my_table}
    \begin{adjustbox}{width=\textwidth}
    \begin{tabular}{l|cccccc|c}
        \toprule
        \textbf{Method} & \textbf{Emotion} & \textbf{Impact} & \textbf{Rhythm} & \textbf{Smoothness} & \textbf{Atmosphere} & \textbf{Surprise} & \textbf{Overall} \\
        \midrule
        \multicolumn{8}{c}{\textbf{ID Evaluation}} \\
        \midrule
        {Ours} 
        & \underline{4.27$\pm$0.98} 
        & \textbf{4.83$\pm$1.02} 
        & \textbf{4.80$\pm$1.04} 
        & \underline{4.47$\pm$1.04} 
        & \textbf{4.40$\pm$0.96} 
        & \textbf{4.48$\pm$1.10} 
        & \textbf{4.54$\pm$0.88} \\

        {Ground Truth} 
        & \textbf{4.46$\pm$1.03} 
        & \underline{4.20$\pm$1.03} 
        & \underline{4.56$\pm$0.90} 
        & \textbf{4.62$\pm$0.81} 
        & \underline{4.32$\pm$0.83} 
        & \underline{4.13$\pm$0.68} 
        & \underline{4.38$\pm$0.74} \\

        {Skip-BART} 
        & 4.06$\pm$0.98 
        & 3.90$\pm$0.91 
        & 4.01$\pm$0.95 
        & 3.91$\pm$1.13 
        & 4.02$\pm$0.97 
        & 3.51$\pm$1.00 
        & 3.90$\pm$0.84 \\

        {Rule-based} 
        & 3.29$\pm$1.39 
        & 2.82$\pm$1.54 
        & 2.43$\pm$1.37 
        & 2.56$\pm$1.26 
        & 2.77$\pm$1.48 
        & 2.36$\pm$1.44 
        & 2.70$\pm$1.26 \\

        \midrule
        \multicolumn{8}{c}{\textbf{OOD Evaluation}} \\
        \midrule
        {Ours} 
        & \textbf{3.72$\pm$1.50} 
        & \textbf{4.36$\pm$1.47} 
        & \textbf{3.96$\pm$1.32} 
        & \textbf{4.08$\pm$1.42} 
        & \textbf{3.86$\pm$1.45} 
        & \textbf{3.66$\pm$1.44} 
        & \textbf{3.94$\pm$1.32} \\

        {Skip-BART} 
        & \underline{3.57$\pm$1.05} 
        & \underline{3.38$\pm$1.03} 
        & \underline{3.69$\pm$1.15} 
        & \underline{3.60$\pm$1.14} 
        & \underline{3.38$\pm$1.12} 
        & \underline{3.19$\pm$1.11} 
        & \underline{3.47$\pm$1.01} \\

        {Rule-based} 
        & 3.06$\pm$1.52 
        & 2.66$\pm$1.52 
        & 2.50$\pm$1.42 
        & 2.47$\pm$1.47 
        & 2.94$\pm$1.61 
        & 2.57$\pm$1.53 
        & 2.70$\pm$1.36 \\

        \bottomrule
    \end{tabular}
    \end{adjustbox}
\end{table*}

\bibliographystyle{ieeetr}
\bibliography{ref.bib}

%%%%%%%%%%%%%%%%%%%%%%%%%%%%%%%%%%%%%%%%%%%%%%%%%%%%%%%%%%%%

\newpage
\appendix

\section*{Appendix Contents}  % 附录目录标题
\startcontents  % 开始记录局部目录
\printcontents{}{1}{\setcounter{tocdepth}{2}}  % 生成目录，深度为section
\newpage

\section{Related Work} \label{sec:related work}
\subsection{Automatic Stage Light Control}

Limited by data scarcity, most early-stage ASSLC research focused on rule-based approaches. These methods first map music into discrete categories, such as chords \cite{mabpa2021automatic}, emotions \cite{bonde2018auditory,hsiao2017methodology,moon2015mood}, or styles \cite{stanescu2018automatic,lei2021music,kanno2022automatic,liao2023automatic}, and then associate each category with a predefined lighting pattern. However, they suffer from several significant shortcomings: (i) \textbf{low interpretability}: predefined lighting patterns often lack empirical justification, and some humanities research questions the strength of the relationship between concepts like emotion and stage lighting \cite{zhao2026automatic,mcdonald2022illuminating}; (ii) \textbf{coarse granularity}: due to the limitations of MIR datasets, most classification modules support only very coarse categories and overlook finer details such as colorful chord textures and music subgenres; (iii) \textbf{low accuracy}: the overall performance of rule-based pipelines is highly sensitive to classification quality, which remains a challenge in MIR. To address some of these issues, prior work used autoencoders \cite{tyroll2020avai} to extract music features and then mapped embeddings to lighting patterns, partially mitigating the granularity and mapping problems.

In contrast, Skip-BART \cite{zhao2026automatic} introduced a different paradigm by learning lighting control directly from professional light engineers and released the RPMC-L$^2$ dataset collected from real livehouse performances. Skip-BART is trained end-to-end on this dataset and achieves performance comparable to human light engineers. However, all the above ASLC works concentrated on generating a single primary light, which limits applicability in real livehouse settings with multiple lights at different positions. More recently, \cite{robinson2026glow} proposed an agent-based system that extract features from video and audio and generate editable ambient light objects, but their system differs from stage-lighting applications. LightGen \cite{wang2026lightinggen} proposed a more direct end-to-end solution that trains a model to map music directly to stage light control parameters (i.e. DMX). However, it introduces a new challenge: different stages have varying numbers of lights and setups, making the trained model difficult to transfer. Moreover, collecting professional data for each stage remains challenging. Even though they collected data in a virtual simulation scenario with expert light engineers, the dataset spans less than three hours, and neither the dataset nor the simulator is publicly available, hindering further research.

In this paper, we propose a novel hierarchical solution: we first train Skip-BART to predict the light hue and value distributions from real-collected stage light videos, and then train a RL framework to decompose the distribution into individual lights. Compared to LightGen, our solution can adaptively transfer to arbitrary stage setups, as the RL process requires no ground-truth annotations and any valid decomposition solution is acceptable.

\subsection{Imitation Learning}
Imitation Learning (IL) \cite{zare2024survey} is a fundamental paradigm for enabling agents to learn from expert demonstrations, with widespread applications in energy management \cite{hu2026offline}, ride hailing \cite{oda2021equilibrium}, and robot control \cite{bing2020energy}. The most straightforward approach within IL is Behavioral Cloning (BC), which frames the problem as supervised learning: the policy is trained to directly predict and replicate expert actions. However, BC suffers from limited generalization; the learned policy can fail catastrophically when encountering states not covered by the offline expert trajectories.

Another major branch of IL is Inverse Reinforcement Learning (IRL) \cite{arora2021survey}, which seeks to infer the underlying reward function (or objective) that expert behaviors are optimizing. By recovering this reward function, IRL can subsequently train a new policy via Reinforcement Learning (RL) that captures the intent behind expert demonstrations, leading to more robust and generalizable performance. A foundational work in this direction is Apprenticeship Learning \cite{abbeel2004apprenticeship}, which assumes the expert's reward function can be represented as a linear combination of known features, and aims to find a policy whose feature expectations match those of the expert. However, this formulation often yields an ambiguous reward function, as multiple reward structures can explain the same expert behavior. This ambiguity arises from the underlying assumption that expert behavior is deterministic and that the demonstrated trajectories are uniquely optimal. To address this limitation, the Maximum Entropy IRL (MaxEnt IRL) framework \cite{ziebart2008maximum} was proposed, which models expert trajectories as being distributed according to the maximum entropy principle subject to feature-matching constraints. This provides a principled way to handle inherent noise and suboptimality in human demonstrations, recovering a probabilistic model of behavior that is both unique and well-calibrated.

In recent decades, the success of adversarial learning, exemplified by Generative Adversarial Networks (GANs) \cite{goodfellow2014generative}, has also transformed the field of IL. A pivotal development was Generative Adversarial Imitation Learning (GAIL) \cite{ho2016generative}, which directly learns a policy that matches the state-action occupancy measure of the expert. GAIL achieves this by training a discriminator to distinguish between expert and generated trajectories, using its output as a surrogate reward signal. This approach bypasses explicit reward function learning and scales effectively to complex, high-dimensional domains. Building on this foundation, the connection between GAIL and IRL was further strengthened by formulations such as Guided Cost Learning (GCL) \cite{finn2016guided}, which integrates IRL within a maximum entropy inverse optimal control objective using energy-based models and optimizes it via sample-based estimation. This line of work culminated in the GAN-GCL framework \cite{finn2016connection}, which explicitly casts the IRL problem as a generative adversarial game, simultaneously learning both a cost function and a policy. Subsequently, Adversarial Inverse Reinforcement Learning (AIRL) \cite{fu2018learning} extended these ideas by introducing a more structured and practical reward formulation that is robust to changes in environment dynamics, disentangling the reward function from the dynamics model and thereby enabling better transfer of the learned reward to new tasks. Note that the goal of this paper is not to develop novel IL methods, but rather to adapt existing IL techniques to address our specific task.
   
\section{Function Definition}  \label{sec:Preliminary}
In this section, we formally define the distance and divergence metrics used throughout this paper, including L1 distance, Wasserstein distance, JS divergence, KL divergence, Bhattacharyya distance, and cosine similarity. Their definitions are as follows:
\begin{align}
&\cdot \text{L1 distance:} \quad D_{\mathrm{L1}}(P \parallel Q) = \sum_{i} |P_i - Q_i|, \label{eq:l1} \\
&\cdot \text{Wasserstein distance:} \quad D_{W}(P \parallel Q) = \inf_{\gamma \in \Pi(P,Q)} \mathbb{E}{(x,y)\sim\gamma} [|x-y|], \label{eq:wasserstein} \\
&\cdot \text{JS divergence:} \quad D_{\mathrm{JS}}(P \parallel Q) = \frac{1}{2} D_{\mathrm{KL}}(P \parallel \frac{P+Q}{2}) + \frac{1}{2} D_{\mathrm{KL}}(Q \parallel \frac{P+Q}{2}), \label{eq:js} \\
&\cdot \text{KL divergence:} \quad D_{\mathrm{KL}}(P \parallel Q) = \sum_{i} P_i \log \frac{P_i}{Q_i}, \label{eq:kl} \\
&\cdot \text{Bhattacharyya distance:} \quad D_{\mathrm{B}}(P \parallel Q) = -\ln \left( \sum_{i} \sqrt{P_i Q_i} \right), \label{eq:bhattacharyya} \\
&\cdot \text{Cosine similarity:} \quad \text{CosSim}(P,Q) = \frac{\sum_{i} P_i Q_i}{\sqrt{\sum_{i} P_i^2} \sqrt{\sum_{i} Q_i^2}}, \label{eq:cosine}
\end{align}
where $P,Q$ are two distributions, $\Pi(P,Q)$ denotes the set of all joint distributions with marginals $P$ and $Q$. For each distribution, we treat it as a discrete vector representing the probability mass in each bin. 

Additionally, for the functions $D_h$ and $D_v$ used in Equation \eqref{eq:restriction-sample} to measure the hue and value distances between two frames, we follow the definitions in \cite{zhao2026automatic}:
\begin{align}
&\cdot \text{Hue distance:} \quad D_{h}(x \parallel y) = \min\{ |x-y|, 2\pi - |x-y|\}, \label{eq:d_h} \\
&\cdot \text{Value distance:} \quad D_{v}(x \parallel y) = |x-y|, \label{eq:d_v} 
\end{align}
where $x$ and $y$ are scalar values.

For the action output in our IL framework, we employ the Von Mises and Beta distributions, whose probability density functions are given by:
\begin{equation}
\mathrm{VonMises}(x \mid \mu, \kappa) \;=\; \frac{\exp\big(\kappa \cos(x-\mu)\big)}{2\pi \mathrm{I}_0(\kappa)} \quad (x \in [0,2\pi)),
\label{eq:von_mises}
\end{equation}
where $\mu$ and $\kappa>0$ (mean direction and concentration) are predicted by the policy network, and $\mathrm{I}_0$ is the modified Bessel function of the first kind of order zero, defined as:
\begin{equation}
\mathrm{I}_0(\kappa) = \frac{1}{\pi} \int_0^{\pi} \exp(\kappa \cos \theta) \, d\theta.
\end{equation}
\begin{equation}
\mathrm{Beta}(x \mid \alpha, \beta) \;=\; \frac{x^{\alpha-1}(1-x)^{\beta-1}}{\mathrm{B}(\alpha,\beta)} \quad (x \in [0,1]),
\label{eq:beta}
\end{equation}
where $\alpha>0,\beta>0$ are shape parameters predicted by the policy, and $\mathrm{B}(\alpha,\beta)=\frac{\Gamma(\alpha)\Gamma(\beta)}{\Gamma(\alpha+\beta)}$ is the Beta function, where
\begin{equation}
\Gamma(\alpha) = \int_0^\infty t^{\alpha-1} e^{-t} \, dt .
\end{equation}

We further define the standard goal-conditioned RL value functions as follows:
\begin{equation}
\begin{aligned}
\mathscr{V}_\pi(s_t,g) & = \mathbb{E}_\pi \left[ \sum_{i=t}^{n} \gamma^{i-t} \mathscr{R}(s_t,a_t,g) \mid s_t \right] , \\
\mathscr{Q}_\pi(s_t, a_t, g) & = \mathbb{E}_\pi \left[ \sum_{i=t}^{n} \gamma^{i-t} \mathscr{R}(s_t,a_t,g) \mid s_t, a_t \right] , \\
\mathscr{A}_\pi(s_t, a_t,g) & = \mathscr{Q}_\pi(s_t, a_t,g) - \mathscr{V}_\pi(s_t,g)  ,
\label{eq:value}
\end{aligned}
\end{equation}
where $\mathscr{V}$, $\mathscr{Q}$, and $\mathscr{A}$ denote the state-value, action-value, and advantage functions, respectively. Specifically, the advantage function in GRPO is defined as:
\begin{equation}
\begin{aligned}
\overline{\mathscr{A}}_{\Phi}(s_{t},a_{t},g) &= \frac{1}{\sigma_{\Phi}^{t:}(g)}\Big( \sum_{i=t}^{n} \gamma^{\,i-t} \mathscr{R}_{\Phi}(s_{i},a_{i},g) - \mu_{\Phi}^{t:}(g) \Big),\\
\mu_{\Phi}^{t:}(g) &= \mathbb{E}_{\pi_{\theta^-}}\Big[ \sum_{i=t}^{n} \gamma^{\,i-t} \mathscr{R}_{\Phi}(s_{i},a_{i},g) \Big],\\
\sigma_{\Phi}^{t:}(g) &= \sqrt{\mathbb{E}_{\pi_{\theta^-}}\Big[ \Big(\sum_{i=t}^{n} \gamma^{\,i-t} \mathscr{R}_{\Phi}(s_{i},a_{i},g) - \mu_{\Phi}^{t:}(g)\Big)^{2} \Big]},
\end{aligned}
\label{eq:grpo_advantage}
\end{equation}
where the states and actions in the first line are computed for a single collected trajectory, while $\mu_{\Phi}^{t:}$ and $\sigma_{\Phi}^{t:}$ are estimated as expectations over multiple trajectories under the behavior policy. (Note that this formulation is a generalized version adapted from the original GRPO used in language modeling \cite{zhao2025one}.) 

\section{Network Architecture} \label{sec:network}
The network architecture for the IL stage is illustrated in Fig. \ref{fig:main}. At time step $t$, the input is the sequence $\mathcal{X}_t = [\langle [\mathrm{SOS}], g \rangle, \langle a_1, \mathrm{Mix}(a_{:1}) \rangle, \langle a_2, \mathrm{Mix}(a_{:2}) \rangle, \ldots, \langle a_{t-1}, \mathrm{Mix}(a_{:t-1}) \rangle]$, where each position $i$ contains the hue-and-value action executed at time $i-1$, i.e., $a_{i-1}$, together with the light distribution aggregated from the actions taken before time $i$, i.e., $\mathrm{Mix}(a_{:i-1})$. For the first position, the distribution is set to the goal $g$, and the action slot is filled with an $[\mathrm{SOS}]$ token to ensure that all positions share the same dimensionality. The sequence $\mathcal{X}_t$ thus serves as a compact representation of $[s_t, g]$. This sequence is fed into a causal Transformer:
\begin{equation}
\mathcal{E}_t = \mathrm{Transformer}(\mathcal{X}_t),
\label{eq:gpt}
\end{equation}
where $\mathcal{E}_t$ is the embedding produced at the final position, which compresses information about the goal and the entire history up to time $t$.

In this work, we adopt an AC framework for the RL component. The embedding $\mathcal{E}_t$ is passed to an actor head and a critic head, each implemented as an MLP. The actor outputs the action parameters (specifically the parameters of the Von Mises distribution $\mu,\kappa$ and the Beta distribution $\alpha,\beta$ defined in the previous subsection), and the critic outputs the state value estimate. To accommodate a learned reward model and the auxiliary objective of predicting state transitions, we introduce three additional MLP heads that take as input the concatenation of $\mathcal{E}_t$ and the chosen action $a_t$. This concatenation serves as a compressed representation of the full tuple $[s_t, a_t, g]$. Concretely, one head predicts the scalar reward $r_t$, while the other two heads predict the resulting hue and value distributions after applying $a_t$.

\section{Experiment Setup} \label{sec:experiment}

\subsection{Dataset Description} \label{subsec:dataset}
% \textbf{\color{red} TODO}

We utilize the publicly available RPMC-L$^2$ dataset \cite{zhao2026automatic}, which comprises recordings from 35 live performances across various commercial venues between December 2020 and July 2024. After data cleaning, videos shorter than 20 seconds were discarded, yielding a total of 699 valid samples. The dataset primarily covers music genres including generalized rock, punk, metal, and core.

For each timestep $t$, we align each music segment with its corresponding video frame $j$ to extract a hue histogram ${h}_j$ and a value histogram ${v}_j$. The histogram counts are normalized so that each frame-wise distribution sums to $1$. For the value histogram, we set the lowest-intensity bin to zero (${v}_j^0 \leftarrow 0$), thereby reducing the contribution of near-black pixels and sensor noise. Through this pre-processing pipeline, we obtain synchronized per-frame distributions ${h}_j$ (360 bins) and ${v}_j$ (100 bins) that serve as the ground truth for our modeling task.

\subsection{Simulation Setup} \label{sec:simulation}

In this paper, we conduct experiments using a simulated multi-light environment rather than real venues, which we leave for future work. \emph{Nevertheless, our method is not limited to simulation and inherently supports real-world scenarios, where the following mixing process could be implemented by collecting data directly from cameras.}

\begin{figure*}[htbp]
\centering
\subfloat[TV ``I Am a Singer'' \cite{singer}]{%
  \label{fig:wsgs}
  \includegraphics[width=0.32\textwidth]{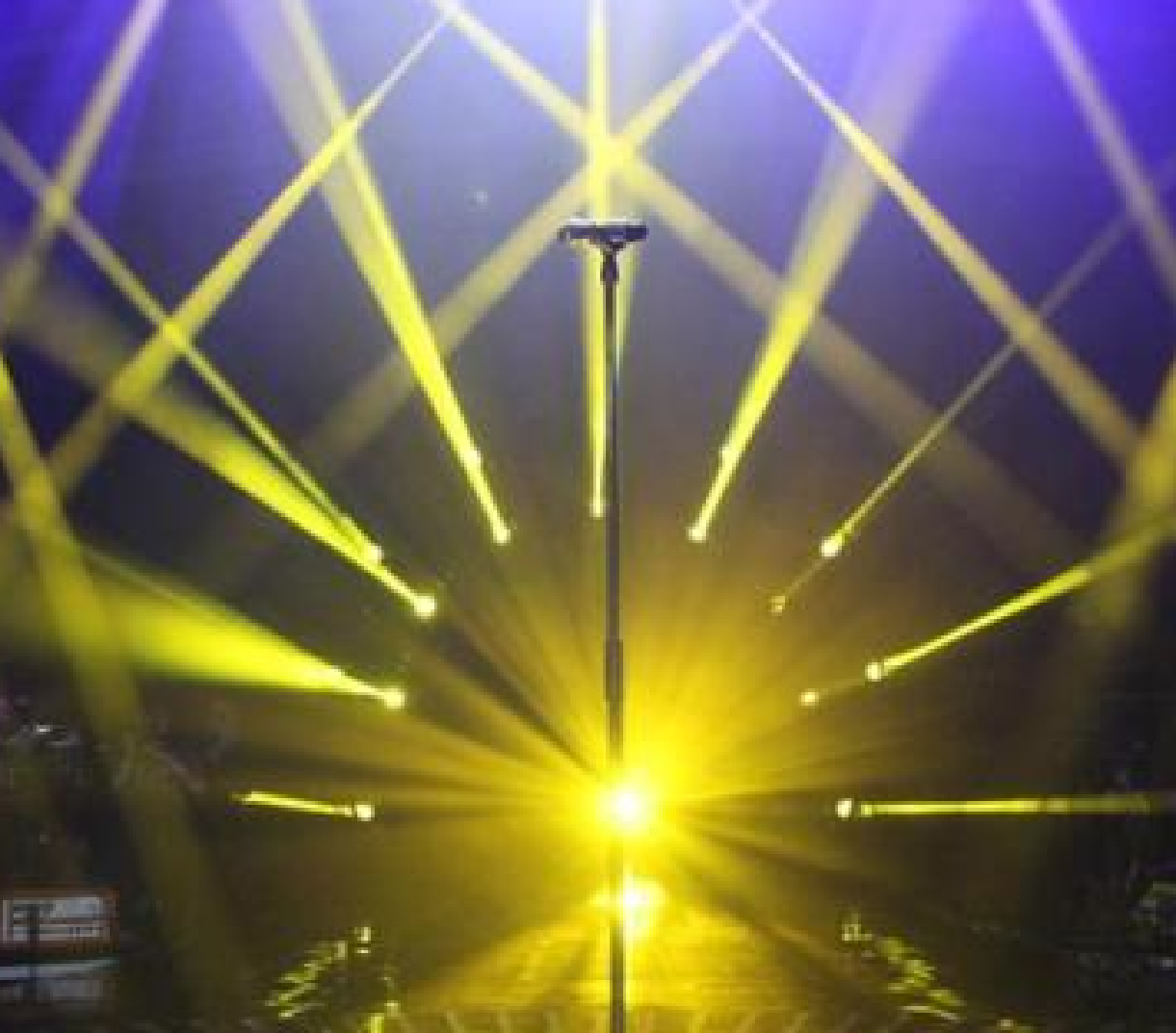}
}
\subfloat[TV ``Sound of My Dream'' \cite{dream}]{%
  \label{fig:mxdsy}
  \includegraphics[width=0.32\textwidth]{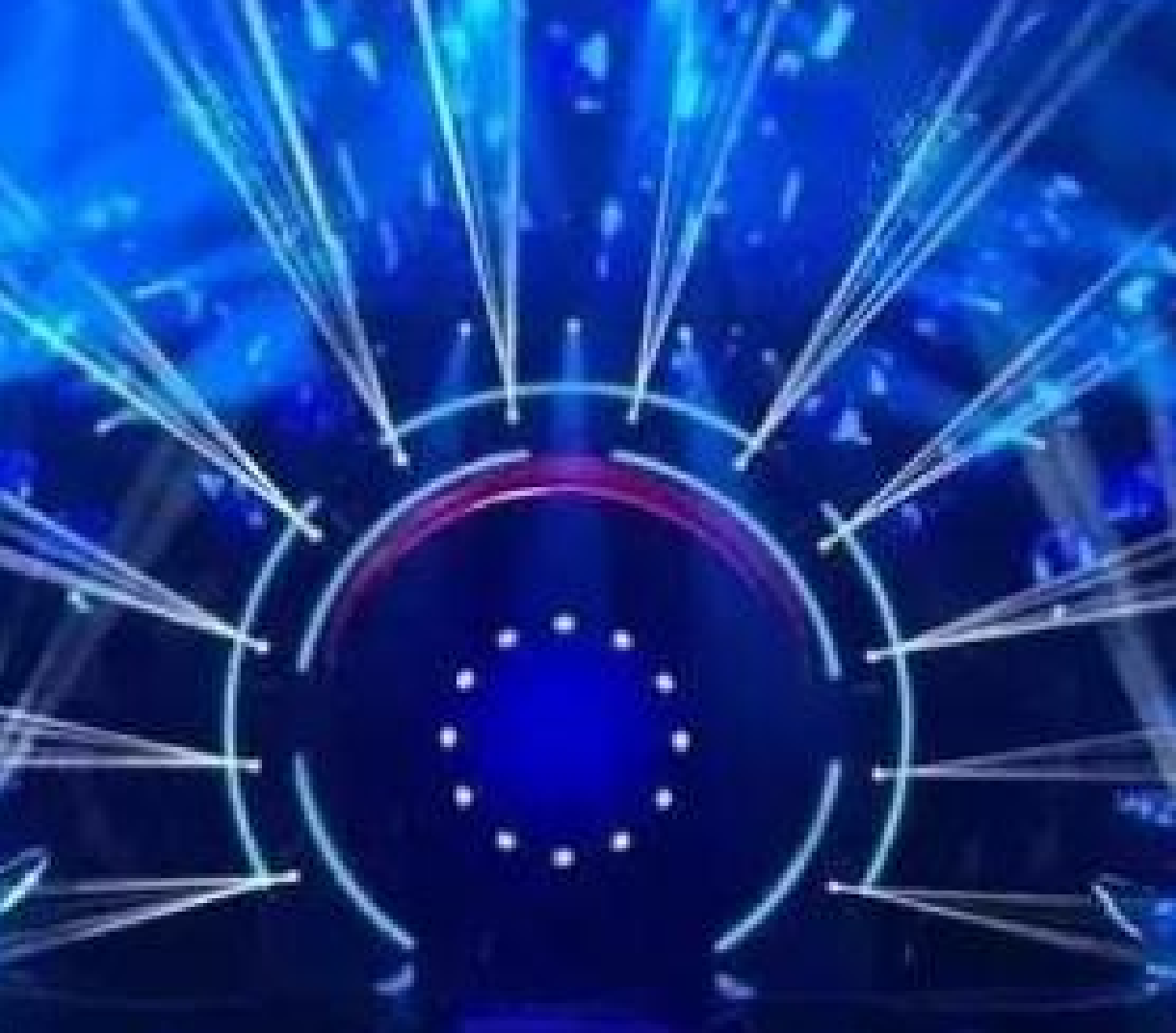}
}
\subfloat[Simulation Scenario]{%
  \label{fig:simulator}
  \includegraphics[width=0.32\textwidth]{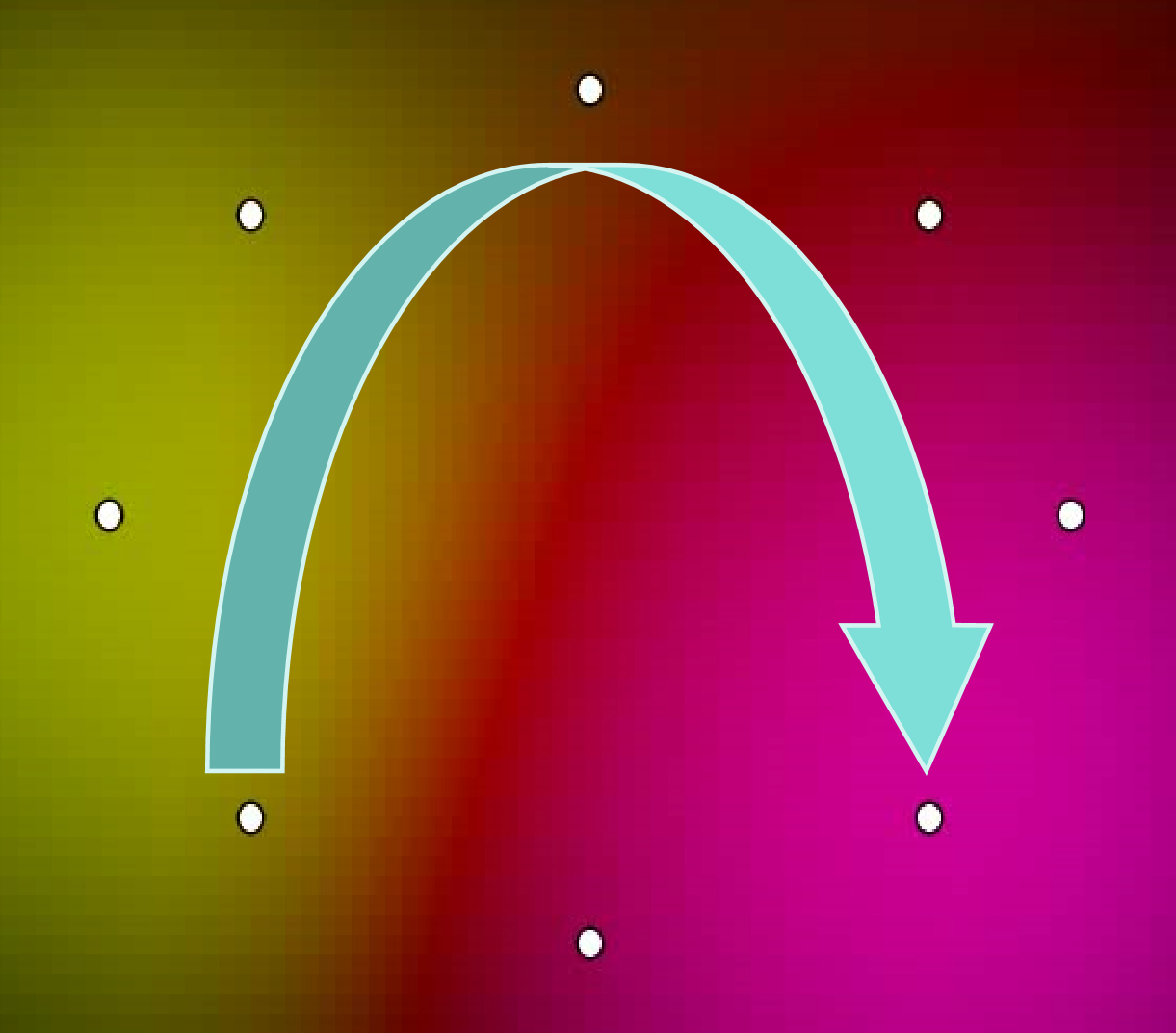}
}
\caption{Examples of Circle Stage Lights: The first two sub-figures are screenshots from TV shows \cite{singer,dream}. In our simulator (the last sub-figure), lights are indexed and actions are generated in clockwise order.}
\label{fig:circle_lights}
\end{figure*}

Specifically, we adopt a circular stage lighting setup, which is commonly used in many television productions, as illustrated in Fig.~\ref{fig:circle_lights}. We consider $N$ lights, each modeled as a point light source, as shown in Fig.~\ref{fig:simulator}. The following describes how we compute the mixed light distribution via simulation, under several simplifying assumptions. Let the image be discretized into a grid of size $H \times W$ (height and width in pixels). For each light $i = 1,\dots,N$, its position in pixel coordinates is $(x_i, y_i)$, where $x_i\in [1,W],y_i \in [1,H]$. The hue $h_i \in [0,2\pi)$ is expressed in radians and the value $v_i \in [0,1]$ is a scalar. \emph{Note that the notation in this section is self-contained and independent of the rest of the paper for clarity.} 

\paragraph{Distance and Weighting \cite{yan2012accurate}:}
For a given pixel at integer coordinates $(u,v)$ with $u=1,\dots,W$, $v=1,\dots,H$, the squared Euclidean distance from light $i$ is
\begin{equation}
d_{i,u,v}^2 = (u - x_i)^2 + (v - y_i)^2.
\label{eq:dist_sq}
\end{equation}
The contribution weight of light $i$ to pixel $(u,v)$ is modeled by a Gaussian decay function: 
\begin{equation}
\begin{aligned}
w_{i,u,v} & = \exp\left( -\frac{d_{i,u,v}^2}{2\sigma_{\text{pix}}^2} \right), \\
\quad \sigma_{\text{pix}} &= \sigma \sqrt{W^2 + H^2},
\label{eq:gaussian}
\end{aligned}
\end{equation}
where $\sigma$ is a relative spread parameter controlling the spatial influence of each light.

\paragraph{Per-Pixel Value \cite{schroeder2000visualizing}:}
The raw value at pixel $(u,v)$ is the weighted sum of the light values:
\begin{equation}
V_{u,v}^{\text{raw}} = \sum_{i=1}^N v_i w_{i,u,v}.
\label{eq:raw_value}
\end{equation}
To prevent excessive brightness, an optional soft clipping may be applied:
\begin{equation}
V_{u,v} = 1 - \exp\big(-cV_{u,v}^{\text{raw}}\big),
\label{eq:clipped_value}
\end{equation}
where $c>0$ is a clipping factor. The final value map is $\{V_{u,v}|u=1,\dots,W; v=1,\dots,H\}$.

\paragraph{Per-Pixel Hue \cite{lei2013vector}:}
The mixed hue at pixel $(u,v)$ is obtained by combining the individual light hues in a vector average. Define the weight for hue mixing as
\begin{equation}
\tilde{w}_{i,u,v} = \frac{v_i w_{i,u,v}}{V_{u,v}^{\text{raw}} + \varepsilon},
\label{eq:hue_weight}
\end{equation}
which ensures that pixels with very low intensity have a negligible hue contribution. The averaged sine and cosine components are
\begin{align}
S_{u,v} &= \sum_{i=1}^N \tilde{w}_{i,u,v} \sin h_i, \label{eq:sin_sum} \\
C_{u,v} &= \sum_{i=1}^N \tilde{w}_{i,u,v} \cos h_i. \label{eq:cos_sum}
\end{align}
Then the mixed hue angle (in radians) is
\begin{equation}
H_{u,v} = \operatorname{atan2}(S_{u,v}, C_{u,v}) \mod 2\pi,
\label{eq:mixed_hue}
\end{equation}
with the convention $H_{u,v}=0$ when $V_{u,v}^{\text{raw}}<\varepsilon$.

\paragraph{Histograms:}
To obtain the overall hue and value distributions, we compute normalized histograms over all pixels. Hue is discretized into $B_h$ bins of equal width $\Delta_h = \frac{2\pi}{B_h}$:
\begin{equation}
\widehat{H}[k] = \frac{1}{HW}\sum_{u,v} \mathbf{1}\left( \left\lfloor \frac{H_{u,v}}{\Delta_h} \right\rfloor = k \right),\quad k=0,\dots,B_h-1,
\label{eq:hue_hist}
\end{equation}
where $\mathbf{1}(\cdot)$ is the indicator function. Value is discretized into $B_v$ bins of width $0.01$:
\begin{equation}
\widehat{V}[\ell] = \frac{1}{HW}\sum_{u,v} \mathbf{1}\left( \lfloor B_v V_{u,v} \rfloor = \ell \right),\quad \ell=0,\dots,B_v-1.
\label{eq:value_hist}
\end{equation}
These histograms serve as the aggregated light distribution $y_j = [\widehat{H}, \widehat{V}]$ in our framework.

\subsection{Model Configurations}

For Skip-BART, we adopt the same configuration as the original paper \cite{zhao2026automatic}, with the exception of modifying the input and output layers for hue and value. Specifically, we replace these layers with MLPs of input and output dimensions 360 and 100, respectively, to accommodate the distributional input and output required by our task. Detailed model configurations are provided in Table \ref{tab:configuration}.

For the IL stage, the configurations are summarized in Table \ref{tab:configuration_il}. Given the well-known difficulty of training large models in RL settings, we adopt a relatively small Transformer architecture compared to Skip-BART.

\begin{table}[htbp]
\centering
\begin{minipage}[t]{0.48\textwidth}
\centering
\caption{Model Configurations of Skip-BART}
\label{tab:configuration}
\begin{tabular}{lc}
\toprule
\textbf{Configuration} & \textbf{Our Setting} \\
\midrule
Bin Amount for Hue and Value  & [360, 100]   \\
Music Embedding & OpenL3 \cite{cramer2019look} \\
Embedding Dimension     & 512   \\
Input Length     & 1024   \\
Number of Network Layers   & 8   \\
Hidden Size   & 2048    \\
Inner Linear Size     & 2048   \\
Attention Heads   & 8   \\
Dropout Rate  & 0.1    \\
Total Number of Parameters  & 231M  \\
Trainable Parameters  & 9 M  \\
\midrule
Optimizer    & AdamW \\
Learning Rate  & 0.0001  \\
Batch Size   & 16  \\
\midrule
Hyper-parameters $[d^h,d^v]$ & [$\frac{\pi}{2}$,0.3] \\
Training Iterations & 200 \\
\bottomrule
\end{tabular}
\end{minipage}%
\hfill
\begin{minipage}[t]{0.48\textwidth}
\centering
\caption{Model Configurations of IL}
\label{tab:configuration_il}
\begin{tabular}{lc}
\toprule
\textbf{Configuration} & \textbf{Our Setting} \\
\midrule
Bin Amount for Hue and Value  & [360, 100]   \\
Embedding Dimension     & 64   \\
Input Length  (Light Amount)   & 8   \\
Number of Network Layers   & 3   \\
Hidden Size   & 64    \\
Inner Linear Size     & 256   \\
Attention Heads   & 4   \\
Dropout Rate  & 0.0    \\
Total Number of Parameters  & 393K  \\
\midrule
Optimizer    & AdamW \\
Learning Rate  & 0.0003  \\
Batch Size   & 64  \\
\midrule
Hyper-parameters $[\eta,\delta,\epsilon]$ & [0.1,0.1,0.2] \\
Phase 1 Iterations & 300 \\
Phase 2 Iterations & 200 \\
Phase 3 Iterations & 500 \\
\bottomrule
\end{tabular}
\end{minipage}
\end{table}

\subsection{Hardware Configurations} \label{sec:hardware}
The experiments were conducted using the PyTorch framework \citep{paszke1912pytorch}. The Skip-BART fine-tuning was performed on a server running Ubuntu 22.04.5 LTS, equipped with an Intel(R) Xeon(R) Gold 6133 CPU @ 2.50 GHz, along with two NVIDIA 4090 GPUs and one NVIDIA A100 GPU. The light decomposition policy was trained concurrently on a separate workstation running Windows 11, equipped with an Intel(R) Core(TM) i7-14700KF processor and an NVIDIA RTX 4080 graphics card.

\section{Quantitative Analysis} \label{sec:Quantitative Analysis}
\subsection{Ablation Study}

In this section, we conduct an ablation study to test model performance when eliminating the BC and AUX losses. The detailed results for hue and value are presented in Tables \ref{tab:hue} and \ref{tab:value}. According to the results, we notice:
\begin{itemize}[left=0pt]
    \item Without BC or AUX loss, model performance worsens in most circumstances, illustrating that they indeed help improve policy performance and enhance the model's capacity to extract features and relationships among input observation sequences. Additionally, when eliminating the BC or AUX loss, performance in Phase 3 decreases compared to Phase 2. This is because the accurate learning of the reward function in AIRL \cite{fu2018learning} relies on the policy achieving optimal performance or being very close to it. When the policy is not sufficiently good, the reward function can be biased, leading the model update in the wrong direction during the Phase 3 fine-tuning.  

    \vspace{0.1cm}
    \item In our task, we observe that PPO does not achieve performance improvement in Phase 3, even with the BC and AUX loss. This is due to the fact that in Phase 2, the critic, actor, and reward model are trained together, which may interfere with each other, making it difficult for the actor to converge and resulting in an incorrect learned reward function. In contrast, GRPO addresses this issue by replacing the advantage estimated by the critic with the group relative reward-to-go.

    \vspace{0.1cm}
    \item Although GRPO achieves the best performance, we still notice that model performance in the ID scenario decreases in Phase 3 compared to Phase 2. This is likely because the model gains high generalization capacity at the expense of overfitting to the original data domain.
\end{itemize}

\begin{table*}[t!]
\centering
\caption{Model Performance on Hue}
\begin{adjustbox}{width=\textwidth}
\begin{tabular}{lcccccc}
\toprule
\multirow{2}{*}{\textbf{Model}} & \multicolumn{2}{c}{\textbf{L1 ($\times 10^{-3}$) $\downarrow$}} & \multicolumn{2}{c}{\textbf{Wasserstein ($\times 10^{-2}$) $\downarrow$}} & \multicolumn{2}{c}{\textbf{JS ($\times 10^{-1}$) $\downarrow$}} \\
\cmidrule(lr){2-3} \cmidrule(lr){4-5} \cmidrule(lr){6-7}
 & \textbf{ID} & \textbf{OOD} & \textbf{ID} & \textbf{OOD} & \textbf{ID} & \textbf{OOD} \\
\midrule
\multicolumn{7}{c}{\textbf{Proposed}} \\
\midrule
Phase 1 & 3.58±1.01 & 3.20±0.76 & 5.61±5.70 & 7.99±0.60 & 3.18±1.25 & 2.80±0.93 \\
Phase 2 (GRPO) & \underline{2.66±0.80} & 2.99±0.59 & 4.97±4.96 & 6.89±0.50 & \underline{2.17±0.93} & 2.71±0.93 \\
Phase 3 (GRPO) & 2.73±0.09 & \textbf{2.59±0.89} & 5.19±5.58 & 7.54±5.75 & 2.23±1.08 & \textbf{2.19±1.11} \\
Phase 2 (PPO) & \textbf{2.52±0.87} & \underline{2.70±0.58} & \textbf{4.73±5.51} & 6.93±5.22 & \textbf{2.03±1.00} & 2.39±0.19 \\
Phase 3 (PPO) & 2.74±0.84 & 3.18±0.53 & \underline{4.80±4.57} & 7.16±4.34 & 2.20±1.02 & 2.87±0.72 \\
\midrule 
\multicolumn{7}{c}{\textbf{w/o BC}} \\
\midrule
Phase 2 (GRPO) & 2.89±0.85 & 3.58±0.40 & 5.06±4.11 & \underline{6.68±4.27} & 2.43±1.02 & 3.29±0.60 \\
Phase 3 (GRPO) & 3.24±1.06 & 3.93±0.36 & 5.75±2.52 & \textbf{6.64±2.93} & 2.80±1.32 &3.19±0.57  \\
Phase 2 (PPO) & 2.79±0.61 & 3.25±0.30 & 5.37±3.07& 6.75±3.37 &2.31±0.81 & 2.88±0.54 \\
Phase 3 (PPO) &2.75±0.92 &2.77±0.45 & 5.90±7.15 & 8.75±6.69&2.84±1.14 & \underline{2.37±0.86}\\
\midrule
\multicolumn{7}{c}{\textbf{w/o AUX}} \\
\midrule
Phase 1 & 2.98±1.03 & 2.84±0.78 & 4.94±5.12 & 7.49±5.53 & 2.56±1.23 & 2.53±0.99 \\
Phase 2 (GRPO) & 2.70±0.86 & 2.97±0.83 & 5.32±5.08 & 7.03±0.51 & 2.25±0.97& 2.60±1.00 \\
Phase 3 (GRPO) & 2.75±0.81& 2.78±0.91 & 5.21±5.22& 8.90±0.82& 2.27±0.92& 2.36±1.03\\
Phase 2 (PPO) & 2.80±0.94& 2.82±0.79 & 5.77±5.17& 8.60±1.11 & 2.30±1.16& 2.42±1.04 \\
Phase 3 (PPO)& 2.98±1.15& 2.89±0.69 &6.76±5.22 & 7.38±0.52 & 2.57±1.39& 2.63±8.15\\
\midrule \midrule
\multirow{2}{*}{\textbf{Model}} & \multicolumn{2}{c}{\textbf{KL $\downarrow$}} & \multicolumn{2}{c}{\textbf{Bhattacharyya ($\times 10^{-1}$) $\downarrow$}} & \multicolumn{2}{c}{\textbf{Cosine ($\times 10^{-1}$) $\uparrow$}} \\
\cmidrule(lr){2-3} \cmidrule(lr){4-5} \cmidrule(lr){6-7}
 & \textbf{ID} & \textbf{OOD} & \textbf{ID} & \textbf{OOD} & \textbf{ID} & \textbf{OOD} \\
\midrule
\multicolumn{7}{c}{\textbf{Proposed}} \\
\midrule
Phase 1 & 1.66±1.56 & 1.15±0.52 & 5.79±2.88 & 4.74±2.00 & 6.64±1.45 & 6.61±1.33 \\
Phase 2 (GRPO) & 3.78±4.64 & 1.98±0.89 & \underline{3.49±1.72} & 4.45±2.13 & 7.13±1.29 & 6.23±1.36 \\
Phase 3 (GRPO) & 1.65±2.33 & \underline{1.05±0.66} & 3.67±2.06 & \textbf{3.61±2.17} & \underline{7.53±1.24} & \textbf{7.19±1.40} \\
Phase 2 (PPO) & 2.58±3.46 & 1.51±5.00 & 3.19±1.87 & \underline{3.87±1.74} & \textbf{7.56±1.21} & \underline{7.18±1.19} \\
Phase 3 (PPO) & 3.77±4.93 & 1.93±6.65 & \textbf{3.05±2.09} & 4.69±1.64 & 7.11±1.27 & 6.42±1.64 \\
\midrule 
\multicolumn{7}{c}{\textbf{w/o BC }} \\
\midrule
Phase 2 (GRPO) & 7.04±0.58 & 2.57±0.48 & 4.01±2.12 & 5.42±1.60& 6.57±1.09& 5.46±0.81 \\
Phase 3 (GRPO) & 9.22±1.58 &2.72±0.36 & 4.89±2.95 & 5.80±1.32& 6.05±1.22 &5.85±1.79  \\
Phase 2 (PPO) & 6.53±5.71& 2.13±0.39  &3.68±1.66 & 4.53±1.38 & 7.02±0.73& 6.30±0.97\\
Phase 3 (PPO) & 1.89±2.95&1.08±3.84 & 3.55±2.23 & 3.75±1.82& 7.18±1.02& 6.27±1.04\\
\midrule
\multicolumn{7}{c}{\textbf{w/o AUX}} \\
\midrule
Phase 1 & \underline{1.28±1.17} & 1.16±0.69 &4.39±2.29 & 4.24±2.07& 7.18±1.47 & 6.85±1.45\\
Phase 2 (GRPO) & 4.40±4.84 & 1.67±0.86 & 3.62±1.89&  4.21±1.95 & 7.22±1.12& 6.59±1.47 \\
Phase 3 (GRPO) & 4.02±4.74& \textbf{0.90±0.39} & 3.68±1.78& 3.89±1.98 & 7.24±1.06& 6.91±1.33 \\
Phase 2 (PPO) &  2.13±2.63 & 1.06±0.50 &3.84±2.40 & 4.01±2.11 & 7.36±1.25& 6.79±1.29 \\
Phase 3 (PPO)& \textbf{1.15±0.73}&1.67±0.78 &4.51±2.91 & 4.25±1.65 &7.20±1.52 & 6.69±1.21 \\
\bottomrule
\end{tabular}
\end{adjustbox}
\label{tab:hue}
\end{table*}

\begin{table*}[t!]
\centering
\caption{Model Performance on Value.}
\begin{adjustbox}{width=\textwidth}
\begin{tabular}{lcccccc}
\toprule
\multirow{2}{*}{\textbf{Model}} & \multicolumn{2}{c}{\textbf{L1 ($\times 10^{-3}$) $\downarrow$}} & \multicolumn{2}{c}{\textbf{Wasserstein ($\times 10^{-2}$) $\downarrow$}} & \multicolumn{2}{c}{\textbf{JS ($\times 10^{-1}$) $\downarrow$}} \\
\cmidrule(lr){2-3} \cmidrule(lr){4-5} \cmidrule(lr){6-7}
 & \textbf{ID} & \textbf{OOD} & \textbf{ID} & \textbf{OOD} & \textbf{ID} & \textbf{OOD} \\
\midrule
\multicolumn{7}{c}{\textbf{Proposed}} \\
\midrule
Phase 1 & 10.21±3.06 & 11.25±1.46 & 5.98±3.38 & 6.87±0.83 & 1.67±0.67 & 2.44±0.43 \\
Phase 2 (GRPO) & \underline{8.63±3.04} & \underline{9.40±1.82} & \underline{5.05±3.32} & \textbf{5.76±1.30} & \underline{1.31±0.65} & \underline{1.87±0.48} \\
Phase 3 (GRPO) & 9.24±3.32 & \textbf{9.14±2.16} & 5.32±3.44 & \underline{5.78±0.79} & 1.50±0.71 & \textbf{1.85±0.53} \\
Phase 2 (PPO) & \textbf{8.07±2.62} & 10.73±2.20 & \textbf{4.74±3.12} & 6.33±1.36 & \textbf{1.21±0.55} & 2.31±6.55 \\
Phase 3 (PPO) & 9.70±3.26 & 11.50±2.53 & 5.60±3.45 & 7.08±1.13 & 1.52±0.70 & 2.57±9.29 \\
\midrule 
\multicolumn{7}{c}{\textbf{w/o BC }} \\
\midrule
Phase 2 (GRPO)  & 9.25±3.25 & 10.86±0.60 & 5.44±3.30& 6.58±0.94 &1.38±0.66 & 2.20±0.51  \\
Phase 3 (GRPO) & 9.56±3.31 & 11.36±1.88& 5.39±3.30 & 6.60±0.94&1.40±0.65 &2.30±0.57 \\
Phase 2 (PPO) & 9.47±3.07 &11.27±1.83 & 5.66±3.02 & 6.87±0.87&1.56±0.73 &2.45±0.57 \\
Phase 3 (PPO) & 10.25±2.98& 12.28±1.71&6.42±3.34 &7.23±1.08 & 1.84±0.68&2.74±0.45 \\
\midrule
\multicolumn{7}{c}{\textbf{w/o AUX}} \\
\midrule
Phase 1 & 9.84±3.15 &11.07±1.67& 5.71±3.39 &6.59±1.03 & 1.62±0.65 & 2.40±0.45\\
Phase 2 (GRPO) &8.96±3.14 & 10.54±2.28 &5.27±3.18 & 6.45±0.92 & 1.40±0.64& 2.26±0.71 \\
Phase 3 (GRPO) & 9.60±2.91& 11.21±2.13 & 5.37±3.25& 6.56±0.83 & 1.54±6.36& 2.47±0.69\\
Phase 2 (PPO) &8.97±3.13 &10.54±2.09 &5.23±3.32 &6.39±0.90 & 1.41±0.65& 2.24±0.60\\
Phase 3 (PPO)&10.47±3.09 & 12.19±1.80 &6.12±3.36 &7.24±0.96 & 1.78±0.68& 2.81±0.62\\
\midrule \midrule
\multirow{2}{*}{\textbf{Model}} & \multicolumn{2}{c}{\textbf{KL ($\times 10^{-1}$) $\downarrow$}} & \multicolumn{2}{c}{\textbf{Bhattacharyya ($\times 10^{-1}$) $\downarrow$}} & \multicolumn{2}{c}{\textbf{Cosine ($\times 10^{-1}$) $\uparrow$}} \\
\cmidrule(lr){2-3} \cmidrule(lr){4-5} \cmidrule(lr){6-7}
 & \textbf{ID} & \textbf{OOD} & \textbf{ID} & \textbf{OOD} & \textbf{ID} & \textbf{OOD} \\
\midrule
\multicolumn{7}{c}{\textbf{Proposed}} \\
\midrule
Phase 1 & 5.76±2.41 & 8.60±1.71 & 2.46±1.14 & 3.78±0.88 & 8.00±0.91 & 6.97±0.60 \\
Phase 2 (GRPO) & \underline{4.54±2.30} & \textbf{6.46±1.85} & 1.85±1.07 & \underline{2.78±0.88} & \underline{8.27±0.86} & \textbf{7.66±0.66} \\
Phase 3 (GRPO) & 5.22±2.49 & \underline{6.55±2.07} & 2.17±1.20 & \textbf{2.70±0.98} & 8.09±0.89 & \underline{7.45±0.74} \\
Phase 2 (PPO) & \textbf{4.44±1.99} & 8.16±2.57 & \textbf{1.64±1.09} & 3.60±1.26 & 8.15±0.75 & 7.13±0.85 \\
Phase 3 (PPO) & 5.24±2.49 & 9.53±4.49 & 2.23±1.17 & 4.28±2.45 & 8.15±0.91 & 6.79±1.24 \\
\midrule 
\multicolumn{7}{c}{\textbf{w/o BC }} \\
\midrule
Phase 2 (GRPO) & 4.81±2.41 & 7.68±2.07 & 1.94±1.06& 3.32±1.23 & 8.26±0.92 & 7.31±0.71   \\
Phase 3 (GRPO) & 4.90±2.42 & 8.14±2.37& 1.96±1.05 & 3.48±1.16& 8.26±0.88 & 7.17±0.78 \\
Phase 2 (PPO) & 5.41±2.44 &8.75±2.31 & 2.26±1.12& 3.85±1.19& 8.06±0.87& 6.91±0.75\\
Phase 3 (PPO) & 6.37±52.52& 9.87±1.91& 2.79±1.20& 4.44±098 & \textbf{8.85±0.99} & 6.53±0.54 \\
\midrule
\multicolumn{7}{c}{\textbf{w/o AUX}} \\
\midrule
Phase 1 & 5.58±2.34 & 8.49±1.82 & 2.31±1.10& 3.72±0.88& 8.04±0.93 & 6.97±0.65 \\
Phase 2 (GRPO) & 4.85±2.28 & 8.04±2.96&2.02±1.05 &3.52±1.41 &8.18±0.87 & 7.15±1.06 \\
Phase 3 (GRPO) & 5.31±2.24 & 8.81±2.96 & 2.24±1.07& 3.92±1.45 &8.09±0.83 & 6.96±0.94 \\
Phase 2 (PPO) & 4.89±2.36 & 7.88±2.45 & 2.02±1.07 & 3.47±1.17 & 8.20±0.88 & 7.23±0.85 \\
Phase 3 (PPO)& 6.11±2.47& 10.25±2.77 & 2.67±1.17& 4.61±1.37 & 7.95±0.97& 6.51±0.85 \\
\bottomrule
\end{tabular}
\end{adjustbox}
\label{tab:value}
\end{table*}

\subsection{RL Training Curves}

\begin{figure*}[htbp]
\centering
\subfloat[GRPO]{%
  \label{fig:grpo}
  \includegraphics[width=0.48\textwidth]{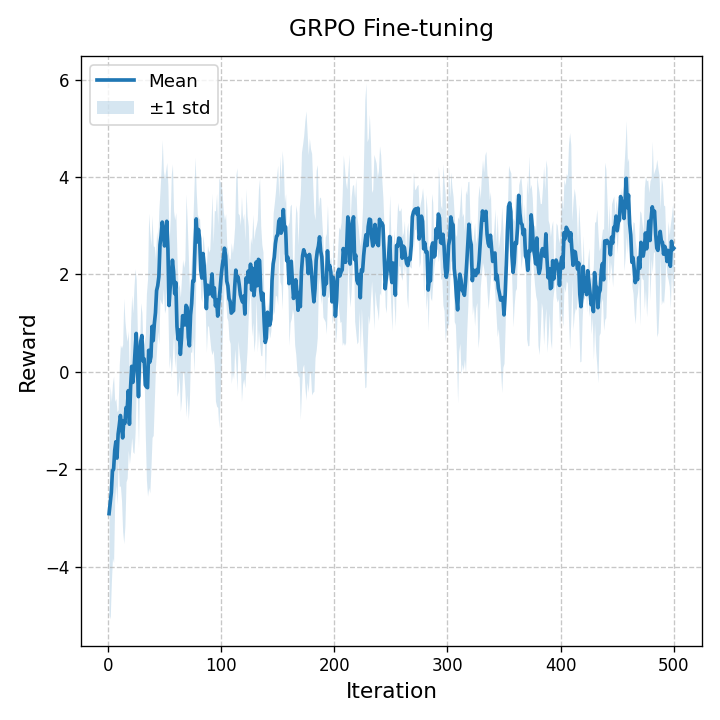}
}
\subfloat[PPO]{%
  \label{fig:ppo}
  \includegraphics[width=0.48\textwidth]{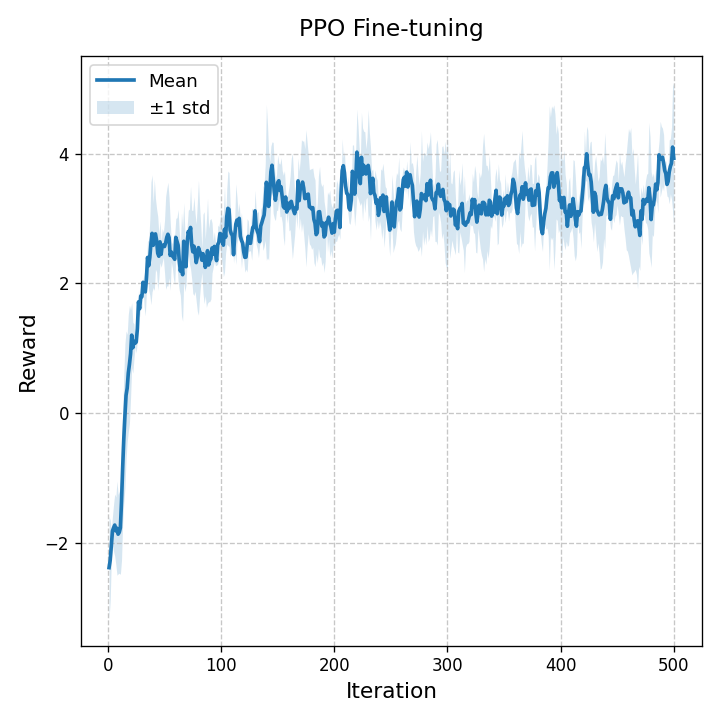}
}
\caption{RL Training Curve in Phase 3.}
\label{fig:convergency}
\end{figure*}

The convergence curve for phase 3 is shown in Fig. \ref{fig:convergency}. Both GRPO and PPO converge in roughly 100 iterations and achieve positive rewards, indicating that they successfully fool the discriminator. Note that because the reward models (discriminators) are trained independently in phase 2 for each method, direct comparison of their absolute reward values is not meaningful.

\section{Human Evaluation} \label{sec:Human Evaluation}

\subsection{Study Setup}

To better assess how our results align with human preferences, we designed a questionnaire to collect participants’ feedback on the stage lighting effects created by each method. We recruited 31 respondents through social media platforms and live music venues, with the human evaluation spanning half a month. After excluding one invalid response (outlier), 30 valid questionnaires were retained for analysis. The participants (based on these 30 valid responses) included 19 males and 11 females aged 18–55 (predominantly 18–30), among whom 7 had professional experience in lighting design, music production, or stage art. All participants reported normal hearing, normal or corrected-to-normal vision, and no history of color blindness or color weakness. All participants are required to evaluate three music pieces alongside the lights across six metrics \cite{erdmann2025development}, with each metric scored from 1 to 7 (the higher the score, the better the evaluation). The lighting conditions were derived from four sources: Ours, Ground Truth, Skip-BART, Rule-based.

Regarding the human evaluation, we follow the same paradigm as Skip-BART \citep{zhao2026automatic}. 
The questionnaire was structured as follows:
\begin{itemize}[left=0pt]
\item \textbf{6 music pieces}: We select six music pieces of different styles to evaluate the generalization capacity of the methods. Specifically, three of them (the ID group) are taken from the testing set of RPMC-L$^2$ and belong to the same domain as the training set, including the styles of rock, music, and core. The other three are generated by Suno \citep{suno} (the OOD group), covering the styles of folk, R\&B, and jazz. Consequently, for the OOD group, there is no ground truth.
\item \textbf{4 objects per group}: Human Light Engineer (HLE), Rule-based, Skip-BART, and proposed SeqLight
\item \textbf{6 evaluation dimensions}: Emotional Match Between Lighting and Music, Visual Impact, Rhythmic Synchronization Accuracy, Smoothness of Lighting Transitions, Immersive Atmosphere Intensity, and Innovative Surprise \citep{erdmann2025development,zhao2026automatic}.
\end{itemize}

\begin{wrapfigure}{r}{0.3\textwidth}
% \begin{figure}[htbp]
\centering 
\includegraphics[width=0.3\textwidth]{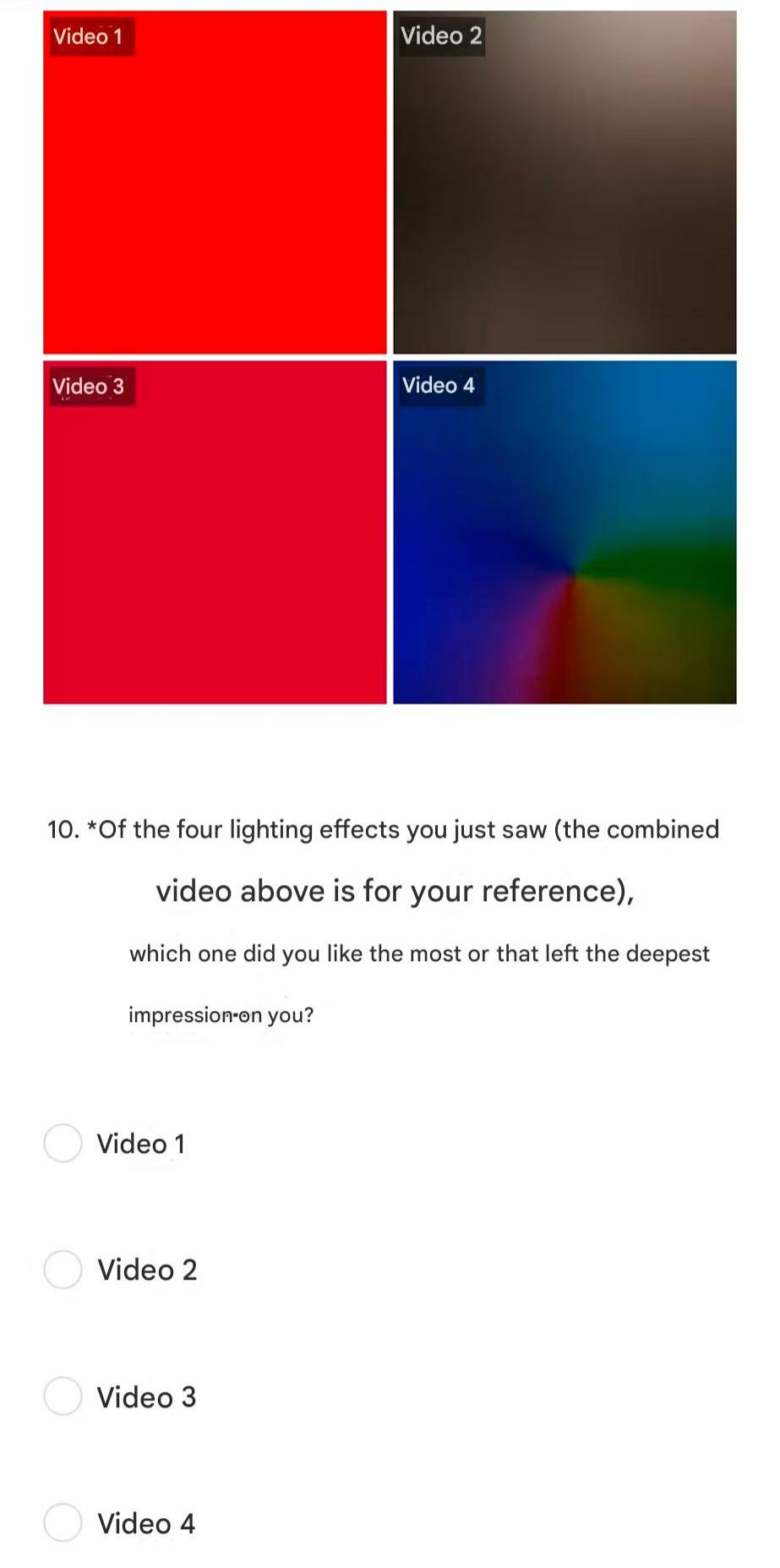}
\caption{The Screenshot of Questionnaire}
\label{fig:question}
% \end{figure}
\end{wrapfigure}

Each music piece formed one group, which included four videos corresponding to the four objects, with the music serving as background sound and a dynamic color block representing the light changes. Participants were asked to rate each video within each music group across six dimensions using a 7-point Likert scale (1 = very dissatisfied, 7 = very satisfied). At the end of each group, participants were also asked to select their favorite video along with a reason (optional), shown as Fig. \ref{fig:question}. More details about the questionnaire can be found in \citep{zhao2024adversarial}.

\subsection{Comparative Objects}
The experimental subjects can be summarized as follows:
\begin{itemize}[left=0pt]
\item \textbf{Human Light Engineer (HLE)}: The HLE is extracted from the original videos in the RPMC-L$^2$ dataset, which consist of live performance footage with lights controlled by professional lighting engineers. For each sample, the corresponding video segment is temporally aligned, resized, and processed with large Gaussian blurring to retain the illumination pattern while suppressing scene details. 
\item \textbf{Rule-Based Method}: The rule-based method first maps music pieces into emotion categories and then maps each category to a predefined light pattern. Following the core idea of \cite{hsiao2017methodology}, we adopt the approach of \cite{alajanki2016benchmarking} to train a bidirectional Long Short-Term Memory (LSTM) model \citep{hochreiter1997long} on the DEAM dataset. The color mapping, which considers both hue and value, is defined according to \cite{nijdam2009mapping}.
\item \textbf{Skip-BART}: Skip-BART \citep{zhao2026automatic} is the first end-to-end stage lighting generation method. It directly maps the music sequence to a single primary color at each frame and is trained on the RPMC-L$^2$ dataset.
\item \textbf{SeqLight}: This refers to our proposed method.
\end{itemize}

% \subsection{Significance Analysis}

% We further conduct statistical significance analysis. In the in-domain setting, our method shows no statistically significant difference from Ground Truth suggesting that the generated results are perceptually comparable to human references. Compared with Skip-BART, our method achieves significant improvements on Impact, Rhythm, Surprise, and Overall quality, while the differences on Emotion, Smoothness, and Atmosphere are not statistically significant. This indicates that our method provides clearer advantages in high-level expressiveness and global perceptual quality.

% In the out-of-domain setting,  our method maintains significant advantages over the rule-based baseline on Impact, Rhythm, Smoothness, Surprise, and Overall quality. Compared with Skip-BART, the improvement is statistically significant on Impact. These results suggest that our method demonstrates stronger robustness and generalization.

\subsection{Results Analysis}

Table~\ref{tab:merged_median_plurality} summarizes the median and plurality scores for ID and OOD evaluations. Our method performs strongly under both criteria, achieving the highest median Overall score in both settings and remaining close to Ground Truth in ID. The plurality results show a similar trend, with our method competitive across most dimensions. These results suggest robust perceptual quality and better generalization than the baselines.

Table~\ref{tab:counts} reports direct preference percentages. In ID, our method receives the highest preference rate (42.22\%), slightly above Ground Truth (40.00\%) and clearly above Skip-BART (16.67\%) and Rule-based (1.11\%). In OOD, our advantage is larger, with 51.11\% preferences compared to 35.55\% for Skip-BART and 13.33\% for Rule-based.

Table~\ref{tab:merged_stats} presents the statistical significance analysis. In ID, our method is not significantly different from Ground Truth, while significantly outperforming Skip-BART on several key dimensions and Rule-based across all dimensions. In OOD, our method significantly improves over Skip-BART on Impact and over Rule-based on most dimensions.

\begin{table*}[htbp]
\centering
\caption{Median and Plurality Scores for both In-domain and Out-of-domain evaluations.}
\label{tab:merged_median_plurality}
% \begin{adjustbox}{width=\textwidth}
\footnotesize
\setlength{\tabcolsep}{3pt}
\begin{tabular}{l*{14}{c}}
\toprule
\multirow{2}{*}{\textbf{Method}} &
\multicolumn{7}{c}{\textbf{ID Evaluation}} &
\multicolumn{7}{c}{\textbf{OOD Evaluation}} \\
\cmidrule(lr){2-8} \cmidrule(lr){9-15}
&
\textbf{Emo} & \textbf{Imp} & \textbf{Rhy} & \textbf{Smo} & \textbf{Atm} & \textbf{Sur} & \textbf{Ovr} &
\textbf{Emo} & \textbf{Imp} & \textbf{Rhy} & \textbf{Smo} & \textbf{Atm} & \textbf{Sur} & \textbf{Ovr} \\
\cmidrule(lr){2-15}
& \multicolumn{14}{c}{\textbf{Median}} \\
\midrule
{Ours}
& 4.00 & \textbf{4.67} & \textbf{4.83} & \underline{4.33} & \textbf{4.33} & \textbf{4.33} & \textbf{4.39}
& \underline{3.50} & \textbf{4.50} & \textbf{4.00} & \textbf{4.00} & \textbf{3.83} & \textbf{4.00} & \textbf{4.08} \\

{Ground Truth}
& \textbf{4.50} & \underline{4.00} & \underline{4.50} & \textbf{4.50} & \textbf{4.33} & \underline{4.00} & \underline{4.31}
& -- & -- & -- & -- & -- & -- & -- \\

{Skip-BART}
& \underline{4.17} & \underline{4.00} & 4.00 & 3.67 & \underline{4.00} & 3.67 & 3.97
& \textbf{3.67} & \underline{3.33} & \underline{3.67} & \underline{3.67} & \underline{3.50} & \underline{3.33} & \underline{3.47} \\

{Rule-based}
& 3.83 & 3.00 & 2.33 & 2.67 & 3.00 & 1.83 & 3.14
& \underline{3.50} & 2.50 & 2.33 & 2.50 & 3.33 & 2.67 & 3.06 \\

\midrule
& \multicolumn{14}{c}{\textbf{Plurality}} \\
\midrule
{Ours}
& \underline{4.00} & \textbf{4.33} & \textbf{5.33} & \underline{3.33} & \underline{4.00} & \textbf{4.00} & \underline{3.94}
& \textbf{3.33} & \textbf{5.00} & \underline{3.00} & \underline{3.00} & \underline{2.67} & \textbf{5.00} & \textbf{2.44} \\

{Ground Truth}
& 3.33 & \underline{3.67} & 3.67 & \textbf{4.33} & \textbf{4.33} & \textbf{4.00} & 3.44
& -- & -- & -- & -- & -- & -- & -- \\

{Skip-BART}
& \textbf{4.67} & \textbf{4.33} & \underline{4.00} & 2.67 & \underline{4.00} & \underline{2.67} & \textbf{4.78}
& \textbf{3.33} & \underline{3.33} & \textbf{4.33} & \textbf{3.67} & \textbf{3.33} & \underline{4.00} & \underline{1.89} \\

{Rule-based}
& \underline{4.00} & 1.00 & 1.00 & 1.00 & 1.00 & 1.00 & 1.00
& \underline{1.33} & 1.00 & 1.00 & 1.00 & 1.00 & 1.00 & 1.06 \\

\bottomrule
\end{tabular}
% \end{adjustbox}
\end{table*}

% 域内和跨域评估计数
% \begin{table}[htbp]
% \centering
% \caption{Evaluation Counts: Which video participants preferred within each group.}
% \label{tab:counts}
% \begin{tabular}{lcc}
% \toprule
% \textbf{Method} & \textbf{ID Evaluation} & \textbf{OOD Evaluation} \\
% \midrule
% {Ours} & \textbf{38} & \textbf{46} \\
% {Ground Truth} & \underline{36} & -- \\
% {Skip-BART} & 15 & \underline{32} \\
% {Rule-based} & 1 & 12 \\
% \bottomrule
% \end{tabular}
% \end{table}

\begin{table}[htbp]
\centering
\caption{Which video participants preferred within each group.}
\label{tab:counts}
\begin{tabular}{lcc}
\toprule
\textbf{Method} & \textbf{ID Evaluation} & \textbf{OOD Evaluation} \\
\midrule
{Ours} & \textbf{42.22\%} & \textbf{51.11\%} \\
{Ground Truth} & \underline{40.00\%} & -- \\
{Skip-BART} & 16.67\% & \underline{35.55\%} \\
{Rule-based} & 1.11\% & 13.33\% \\
\bottomrule
\end{tabular}
\end{table}

\begin{table*}[t!]
\centering
\caption{Statistical comparisons for In-domain and Out-of-domain evaluations. Mean difference ($\Delta$M), standard deviation (SD), and $p$-values. Significance: *** $p<0.001$, ** $p<0.01$, * $p<0.05$. Missing comparisons (due to no Ground Truth in OOD) are marked with `--`.}
\label{tab:merged_stats}
% \begin{adjustbox}{width=\textwidth}
\footnotesize
\setlength{\tabcolsep}{4pt}
\begin{tabular}{l l *{3}{c} *{3}{c}}
\toprule
\multirow{2}{*}{\textbf{Metrics}} & \multirow{2}{*}{\textbf{Comparison}} & 
\multicolumn{3}{c}{\textbf{ID Evaluation}} & \multicolumn{3}{c}{\textbf{OOD Evaluation}} \\
\cmidrule(lr){3-5} \cmidrule(lr){6-8}
& & \textbf{$\Delta$M} & \textbf{SD} & \textbf{$p$} & \textbf{$\Delta$M} & \textbf{SD} & \textbf{$p$} \\
\midrule

\multirow{6}{*}{Emotion} 
& ours vs GT & -0.19 & 0.23 & 1.000 & -- & -- & -- \\
& ours vs SB & 0.21 & 0.21 & 1.000 & 0.16 & 0.30 & 1.000 \\
& ours vs RB & 0.98 & 0.29 & $0.013^{*}$ & 0.67 & 0.37 & 0.239 \\
& GT vs SB   & 0.40 & 0.17 & 0.144 & -- & -- & -- \\
& GT vs RB   & 1.17 & 0.26 & $0.001^{**}$ & -- & -- & -- \\
& SB vs RB   & 0.77 & 0.21 & $0.006^{**}$ & 0.51 & 0.24 & 0.130 \\

\midrule
\multirow{6}{*}{Impact} 
& ours vs GT & 0.63 & 0.23 & 0.053 & -- & -- & -- \\
& ours vs SB & 0.93 & 0.23 & $0.002^{**}$ & 0.98 & 0.28 & $0.005^{**}$ \\
& ours vs RB & 2.01 & 0.32 & $<0.001^{***}$ & 1.70 & 0.37 & $<0.001^{***}$ \\
& GT vs SB   & 0.30 & 0.17 & 0.531 & -- & -- & -- \\
& GT vs RB   & 1.38 & 0.29 & $<0.001^{***}$ & -- & -- & -- \\
& SB vs RB   & 1.08 & 0.22 & $<0.001^{***}$ & 0.72 & 0.23 & $0.011^{*}$ \\

\midrule
\multirow{6}{*}{Rhythm} 
& ours vs GT & 0.24 & 0.22 & 1.000 & -- & -- & -- \\
& ours vs SB & 0.79 & 0.22 & $0.007^{**}$ & 0.27 & 0.27 & 0.989 \\
& ours vs RB & 2.37 & 0.30 & $<0.001^{***}$ & 1.46 & 0.31 & $<0.001^{***}$ \\
& GT vs SB   & 0.54 & 0.20 & 0.058 & -- & -- & -- \\
& GT vs RB   & 2.12 & 0.31 & $<0.001^{***}$ & -- & -- & -- \\
& SB vs RB   & 1.58 & 0.24 & $<0.001^{***}$ & 1.19 & 0.24 & $<0.001^{***}$ \\

\midrule
\multirow{6}{*}{Smoothness} 
& ours vs GT & -0.16 & 0.22 & 1.000 & -- & -- & -- \\
& ours vs SB & 0.56 & 0.22 & 0.103 & 0.48 & 0.26 & 0.218 \\
& ours vs RB & 1.91 & 0.29 & $<0.001^{***}$ & 1.61 & 0.30 & $<0.001^{***}$ \\
& GT vs SB   & 0.71 & 0.23 & $0.030^{*}$ & -- & -- & -- \\
& GT vs RB   & 2.07 & 0.27 & $<0.001^{***}$ & -- & -- & -- \\
& SB vs RB   & 1.36 & 0.21 & $<0.001^{***}$ & 1.13 & 0.22 & $<0.001^{***}$ \\

\midrule
\multirow{6}{*}{Atmosphere} 
& ours vs GT & 0.08 & 0.20 & 1.000 & -- & -- & -- \\
& ours vs SB & 0.38 & 0.20 & 0.424 & 0.48 & 0.28 & 0.284 \\
& ours vs RB & 1.63 & 0.32 & $<0.001^{***}$ & 0.91 & 0.40 & 0.091 \\
& GT vs SB   & 0.30 & 0.17 & 0.571 & -- & -- & -- \\
& GT vs RB   & 1.56 & 0.27 & $<0.001^{***}$ & -- & -- & -- \\
& SB vs RB   & 1.26 & 0.25 & $<0.001^{***}$ & 0.43 & 0.26 & 0.299 \\

\midrule
\multirow{6}{*}{Surprise} 
& ours vs GT & 0.34 & 0.21 & 0.699 & -- & -- & -- \\
& ours vs SB & 0.97 & 0.24 & $0.002^{**}$ & 0.47 & 0.29 & 0.346 \\
& ours vs RB & 2.12 & 0.31 & $<0.001^{***}$ & 1.09 & 0.35 & $0.012^{*}$ \\
& GT vs SB   & 0.62 & 0.17 & $0.006^{**}$ & -- & -- & -- \\
& GT vs RB   & 1.78 & 0.28 & $<0.001^{***}$ & -- & -- & -- \\
& SB vs RB   & 1.16 & 0.27 & $0.001^{**}$ & 0.62 & 0.18 & $0.004^{**}$ \\

\midrule
\multirow{6}{*}{Overall} 
& ours vs GT & 0.16 & 0.19 & 1.000 & -- & -- & -- \\
& ours vs SB & 0.64 & 0.18 & $0.008^{**}$ & 0.47 & 0.26 & 0.226 \\
& ours vs RB & 1.84 & 0.27 & $<0.001^{***}$ & 1.24 & 0.31 & $0.001^{**}$ \\
& GT vs SB   & 0.48 & 0.15 & $0.022^{*}$ & -- & -- & -- \\
& GT vs RB   & 1.68 & 0.24 & $<0.001^{***}$ & -- & -- & -- \\
& SB vs RB   & 1.20 & 0.20 & $<0.001^{***}$ & 0.77 & 0.18 & $0.001^{**}$ \\

\bottomrule
\end{tabular}
% \end{adjustbox}
\end{table*}

\section{Visualizations}

\subsection{Real-World Livehouse Example} \label{sec:mekader}
\begin{figure*}[t!]
\centering
\subfloat[IMPACT MOTION Livehouse, Chongqing, China, 2021]{%
  \includegraphics[width=0.4\textwidth]{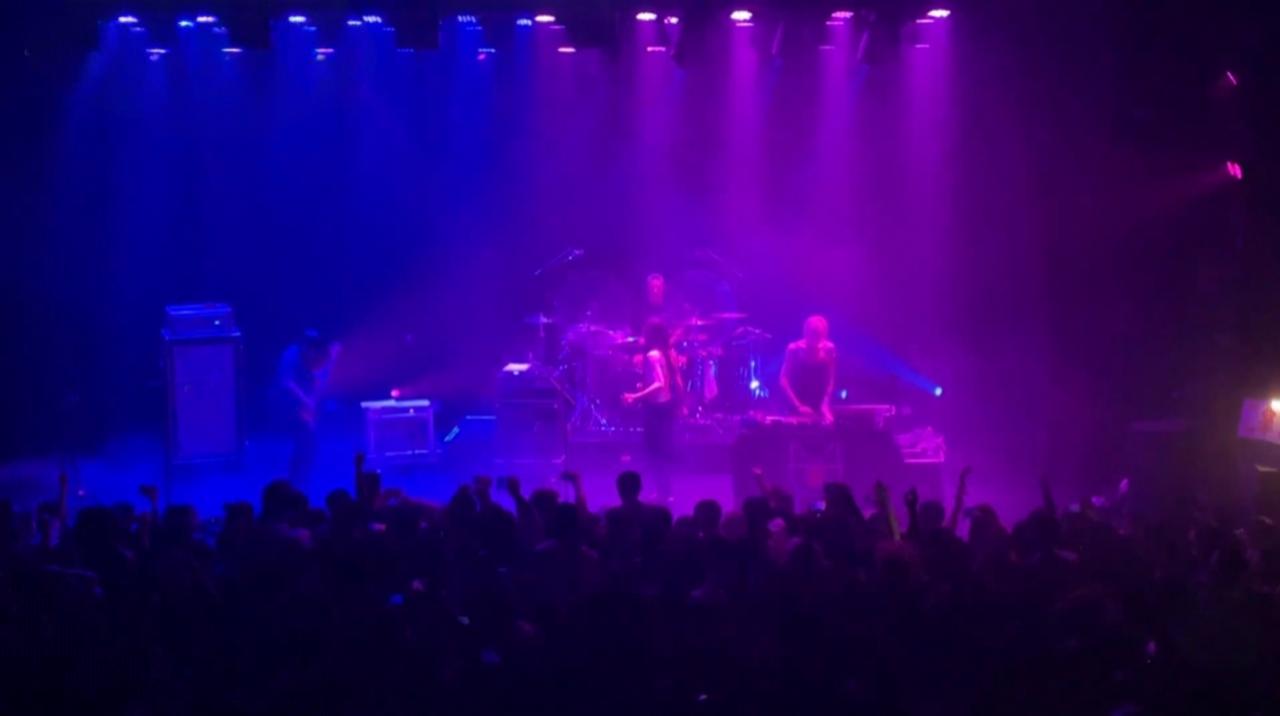}
}
\subfloat[ALSOLIVE Livehouse, Foshan, Guangdong, China, 2025]{%
  \includegraphics[width=0.4\textwidth]{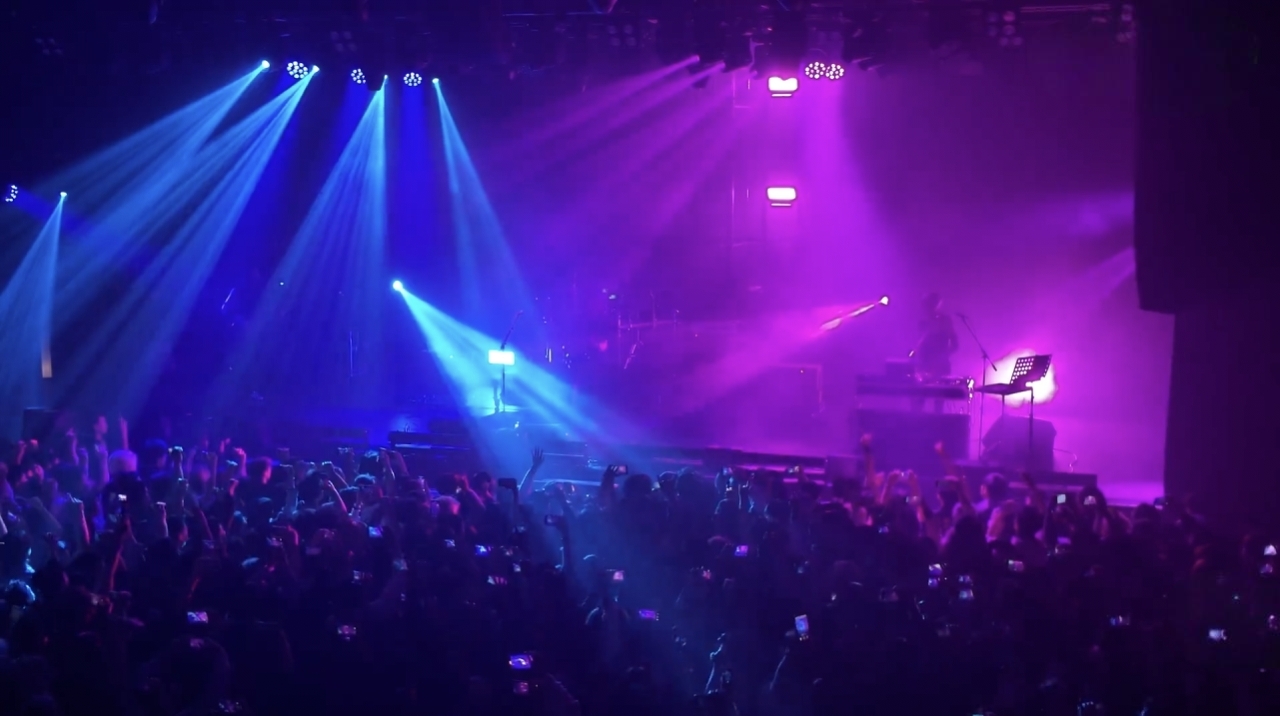}
} \\
\subfloat[HV Distribution (computed from a)]{%
  \includegraphics[width=0.4\textwidth]{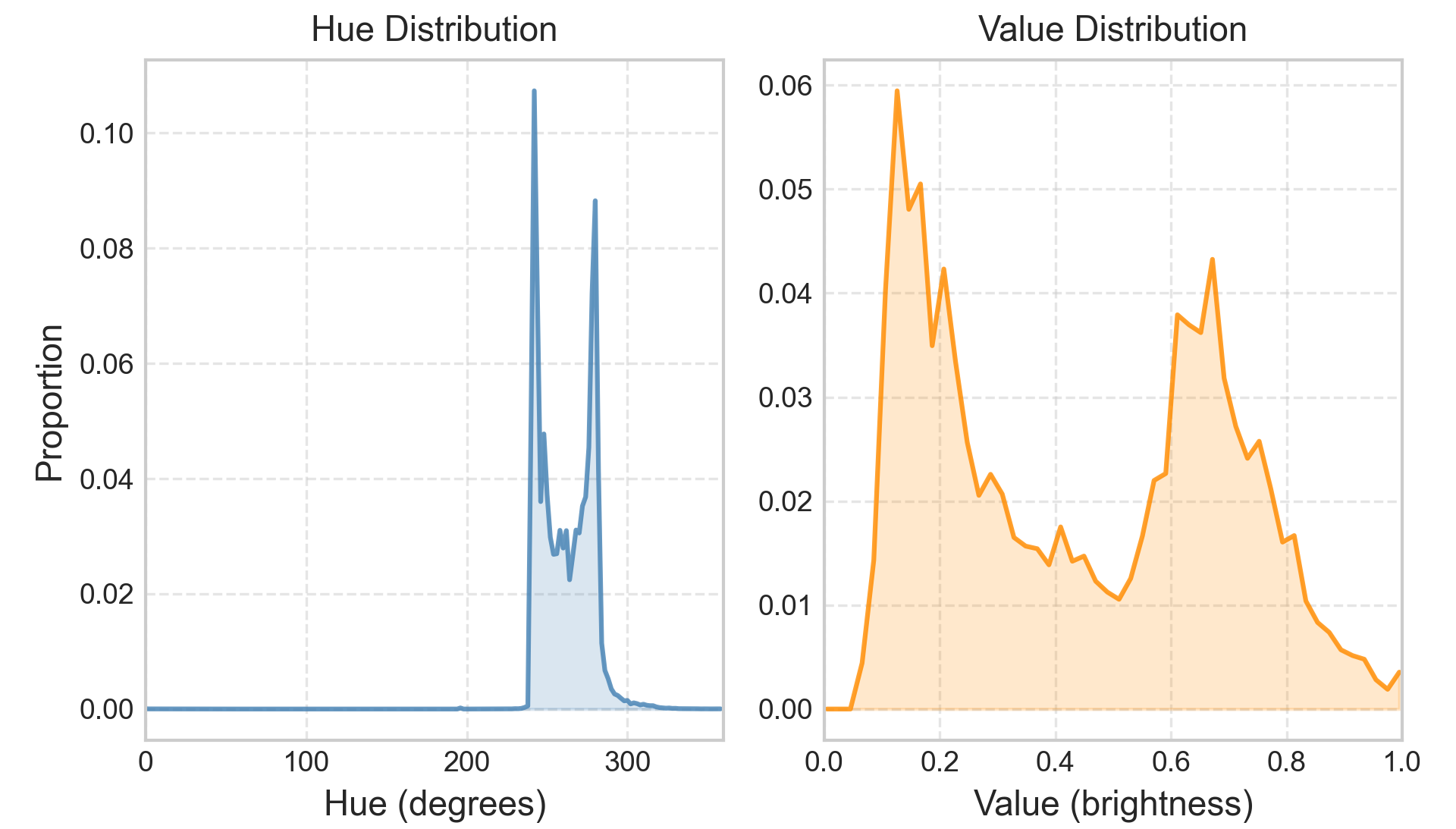}
}
\subfloat[HV Distribution (computed from b)]{%
  \includegraphics[width=0.4\textwidth]{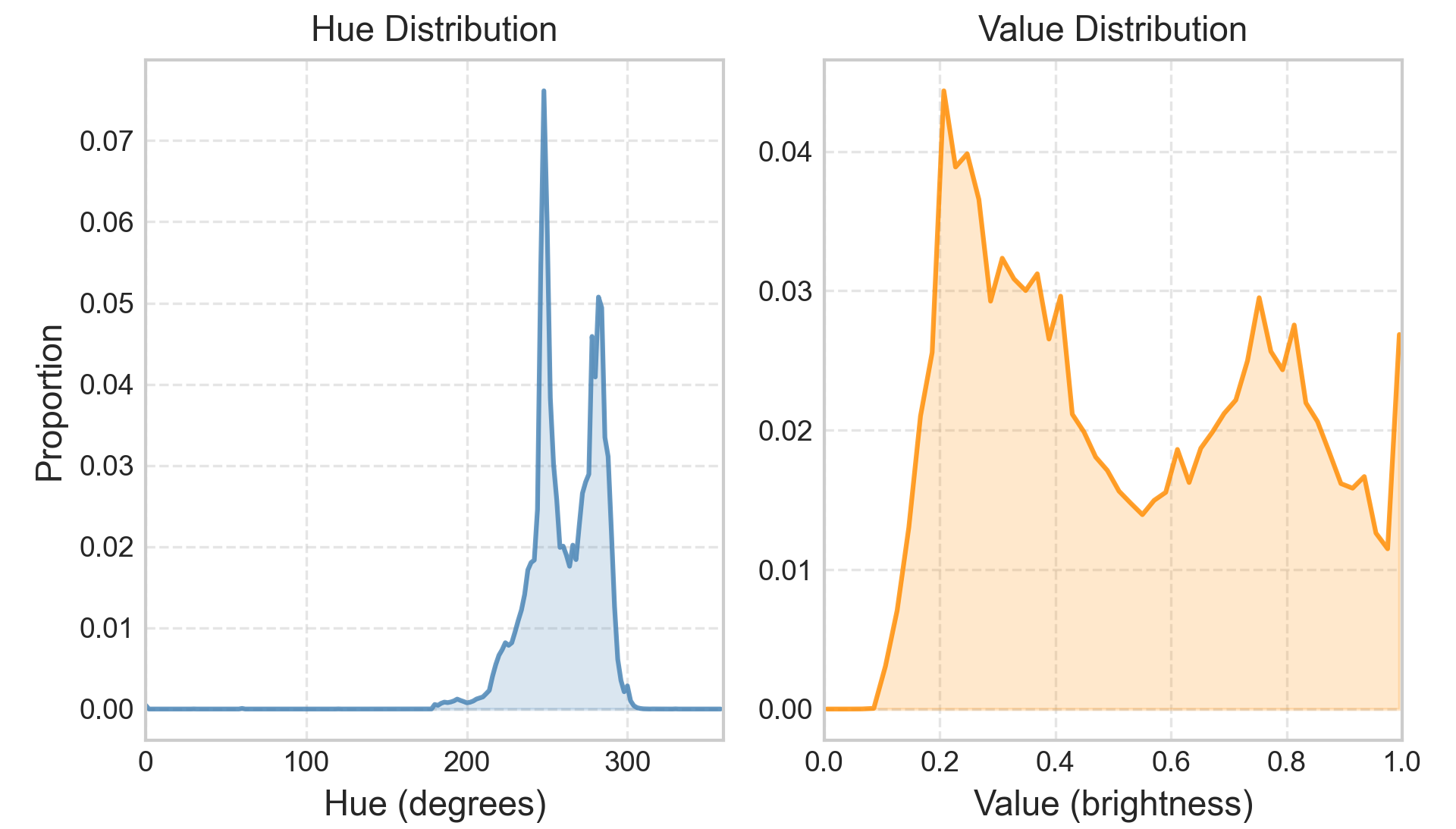}
}
\caption{The Live Performance of Chinese Rock Band Mekader.}
\label{fig:mekader_live}
\end{figure*}

Figure \ref{fig:mekader_live} shows recordings of an intro by the Chinese rock band Mekader from two different performances, in different years and venues. Despite differing lighting setups and recording viewpoints, the designed light patterns remain similar, and the resulting HV distribution shapes are largely consistent. This empirical evidence supports our use of recorded videos from multiple venues to predict the full light distribution, thereby helping to address the data scarcity challenge that has long plagued many MIR fields. (ii) In the second stage, we aim to find an effective way to decompose the distribution obtained from stage 1 into individual light controls. To this end, we formulate the light decomposition task as a GCMDP, decoupling it from any music-related information. Consequently, this stage requires no professional lighting engineers for data collection. Specifically, we propose to solve the GCMDP using IL, where expert trajectories can be readily collected within each venue using only mixed light data (whether from simulation or real image capture). This also means that each venue can independently train its own light decomposition model and combine it with the light distribution prediction model from stage 1. In this way, our method achieves high transferability and practical applicability.

\subsection{Goal-Conditioned Light Decomposition}  \label{sec:visual}

\begin{figure*}[htbp]
\centering
\subfloat[Case 1 Histogram]{%
\includegraphics[width=0.49\textwidth]{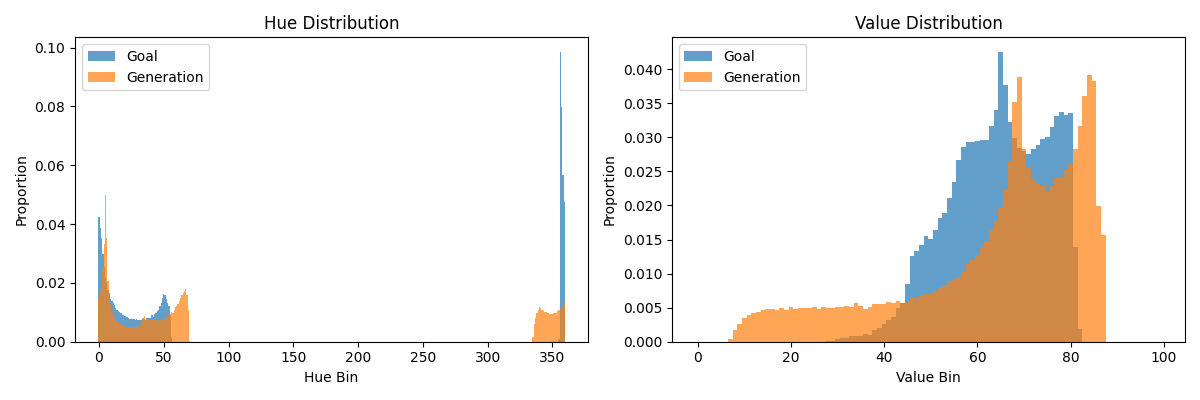}
}
\subfloat[Case 2 Histogram]{%
\includegraphics[width=0.49\textwidth]{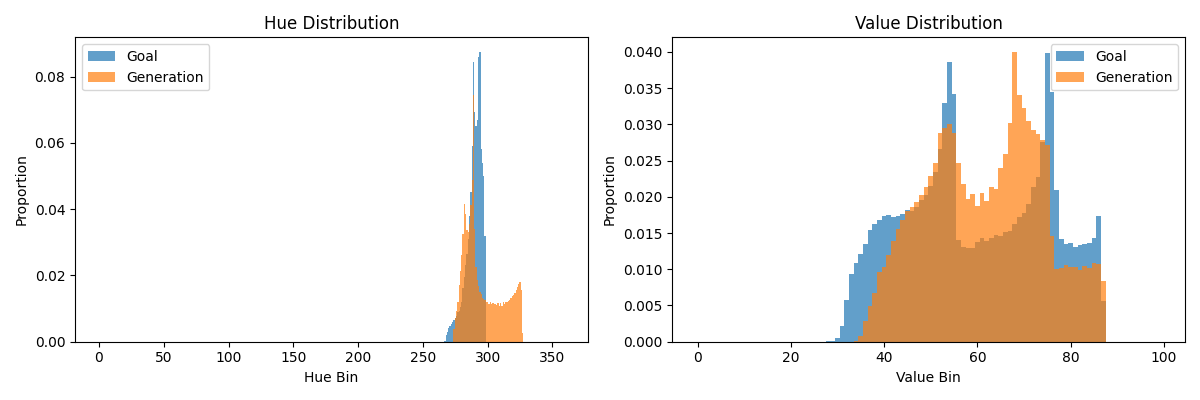}
} \\
\subfloat[Case 3 Histogram]{%
\includegraphics[width=0.49\textwidth]{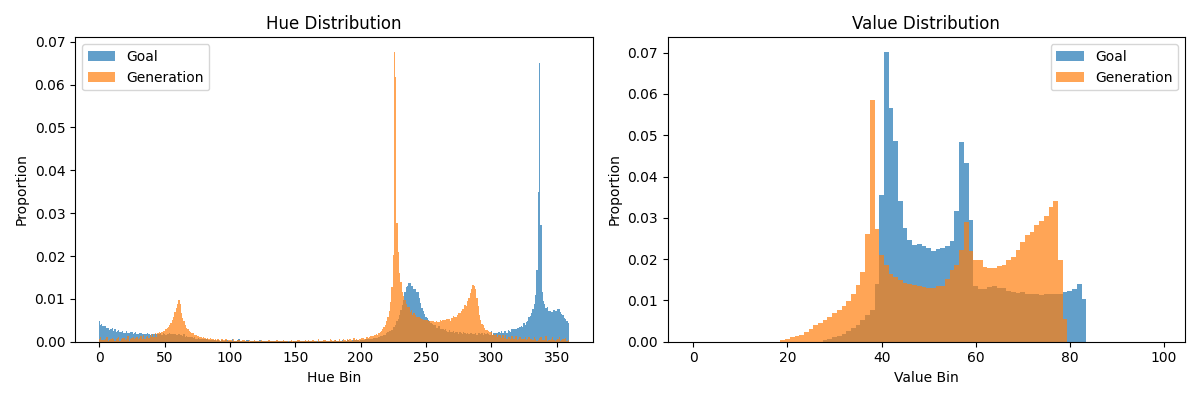}
}
\subfloat[Case 4 Histogram]{%
\includegraphics[width=0.49\textwidth]{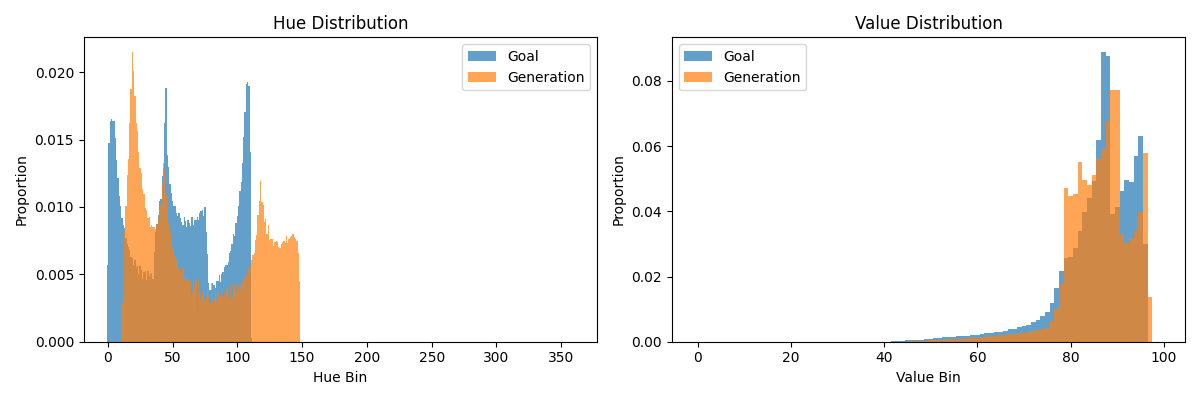}
}\\
\subfloat[Case 1 Light]{%
\includegraphics[width=0.49\textwidth]{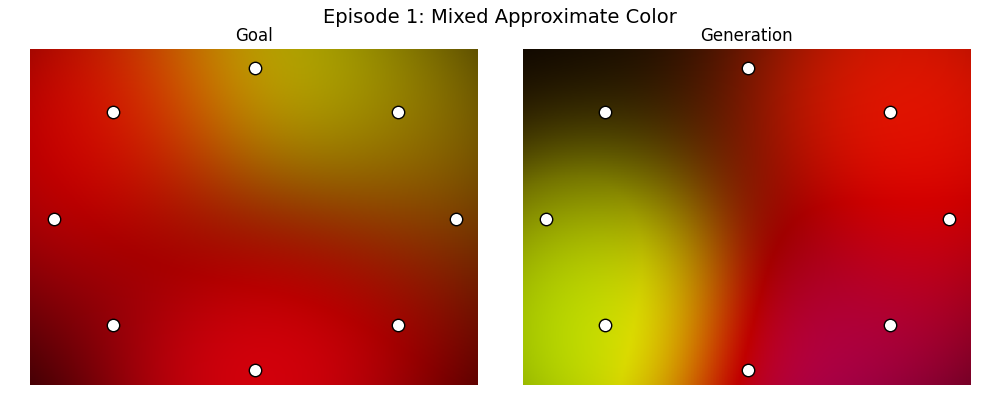}
}
\subfloat[Case 2 Light]{%
\includegraphics[width=0.49\textwidth]{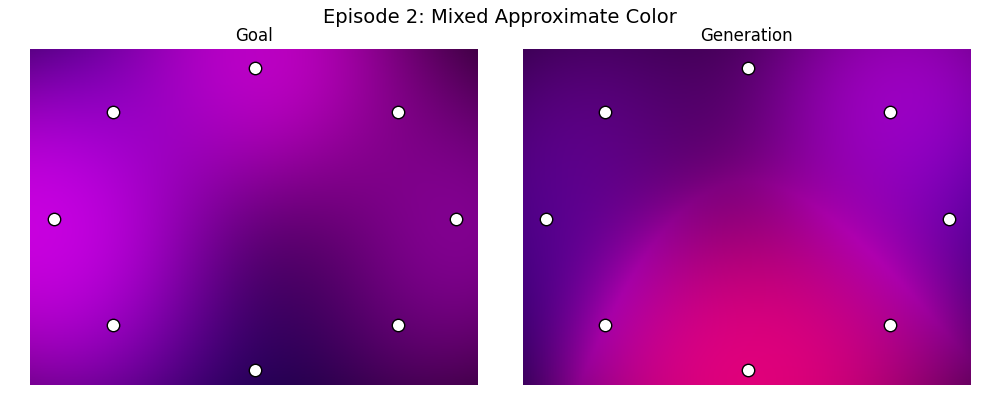}
} \\
\subfloat[Case 3 Light]{%
\includegraphics[width=0.49\textwidth]{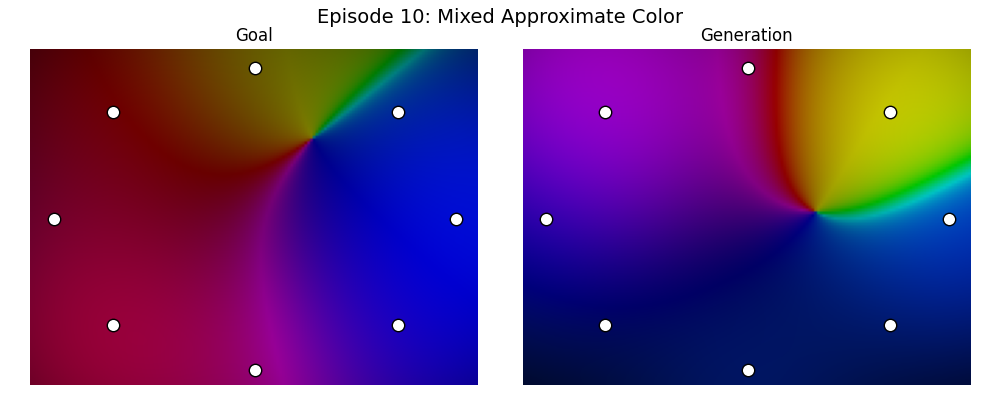}
}
\subfloat[Case 4 Light]{%
\includegraphics[width=0.49\textwidth]{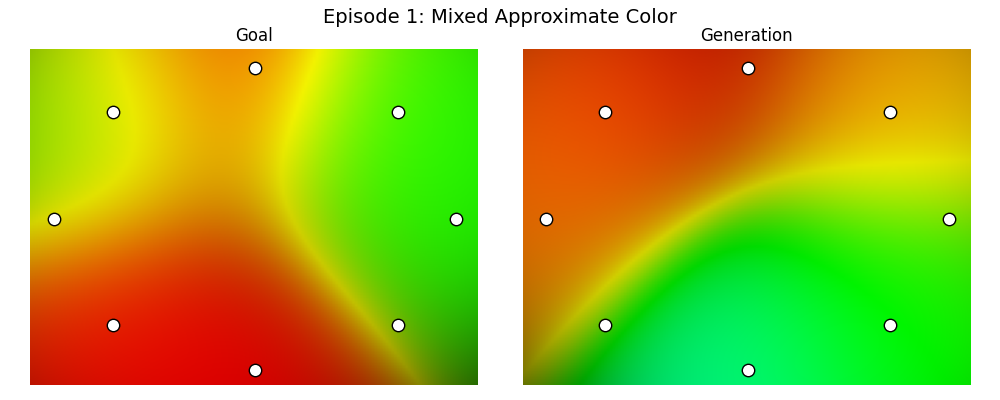}
} 
\caption{Visualization results of goal-conditioned light decomposition. In the histogram plots, blue regions represent the target goal distribution, while orange regions show the distribution produced by our method. For each light case, the left image depicts the ground-truth goal and the right image shows the generated result.}
\label{fig:visual_goal}
\end{figure*}

In this section, we present visualizations of our method applied to the goal-conditioned light decomposition task. Goals are first constructed using our HER-based labeling method, and then decomposed into individual light controls via the proposed imitation learning framework. Representative results are shown in Fig.~\ref{fig:visual_goal}, where each goal comprises either a single color or a combination of multiple colors.

As illustrated in the histogram plots, the generated distributions closely match the target goal distributions. However, due to the inherent ambiguity of the decomposition task, where multiple per-light configurations can yield the same aggregated distribution, the spatial arrangement of the generated lights may differ from that of the ground-truth goal. This positional discrepancy does not adversely affect performance, as our method incorporates each light's previous frame state as a constraint to ensure temporal smoothness and practical control stability. Consequently, even when the generated light positions deviate from those in the goal, the resulting control sequence remains both valid and coherent.

% \subsection{Music-Inspired Multi-Light Generation}

% \textcolor{blue}{\textbf{TODO: add an example}}

\section{Discussions} \label{sec:Discussions}

In this section, we discuss the potential limitations of this work and outline promising directions for future research. Although this paper presents the first color-space multi-light ASLC method, several simplifying assumptions are made. For instance, we assume all lights are point sources, thereby ignoring their directional properties. Moreover, in our experiments, we only consider a simple simulation setup with eight point lights, which simplifies the spatial relationships among lights. Future work could aim to realize this approach in real livehouse venues by utilizing cameras for data collection.

From a technical perspective, several avenues for future research emerge. First, the light direction and cross-frame temporal relationships could be modeled more explicitly. For example, incorporating light direction into the action space could enhance realism, although this would significantly increase the complexity of the problem. Second, we currently perform light decomposition independently for each frame. While we enforce temporal consistency through our proposed constrained sampling strategy, there remains room for improvement. One promising direction is to formulate the task as a Multi-Agent Reinforcement Learning (MARL) problem, treating each light as an independent agent. However, this introduces new challenges, such as achieving effective cooperation among agents, and may require incorporating music information into the state space. Although such an approach could potentially improve the alignment between music and lighting effects, it would also substantially increase training difficulty and computational cost.

Additionally, most existing ASLC methods, including ours, do not support online control due to high computational demands and the requirement of processing the entire music sequence as input. While online control remains an important direction for future investigation, we argue that offline control still holds significant practical value. For instance, many live performances rely on pre-programmed lighting and VJ setups, where a manual trigger or click track is necessary to maintain synchronism between the artists' live performance and the pre-computed elements. In such contexts, pre-computed lighting control is entirely acceptable.

Finally, although our method outputs lighting control in a color space (HV) that can adapt to various venues, a human operator is still required to convert these per-light color values into low-level lighting control parameters (e.g., DMX signals). Future research could explore fully automating this conversion process, thereby further reducing manual intervention.

SeqLight may lower the cost and expertise barrier for music-conditioned stage lighting, making lighting design more accessible to small venues, independent artists, and educational performances. It may also help professional lighting engineers prototype lighting effects more efficiently. Potential negative impacts include reduced demand for some manual lighting design labor.

\end{document}